\documentclass[aps,twocolumn,superscriptaddress,nofootinbib,longbibliography]{revtex4-2}

\usepackage{amsmath}
\usepackage{amssymb}
\usepackage{graphicx}
\usepackage{bm}
\usepackage{xcolor}

\usepackage{url}
\usepackage{array} 
\graphicspath{{figures/}}

\begin{document}

\title{Broken-Symmetry Reference States for Sample-Based Quantum Diagonalization\\
with Near-DMRG Accuracy}

\author{Kunal Kumar}\thanks{Corresponding authors.}
\affiliation{IBM Quantum, IBM Japan, Tokyo 103-8510, Japan}
\email{kunalkumar@ibm.com}

\author{Tomonori Shirakawa}
\affiliation{RIKEN Center for Computational Science (R-CCS), Kobe 650-0047, Japan}
\affiliation{Computational Condensed Matter Physics Laboratory, RIKEN Pioneering Research Institute (PRI), Wako 351-0198, Japan}

\affiliation{RIKEN Center for Quantum Computing (RQC), Wako 351-0198, Japan}
\affiliation{RIKEN Center for Interdisciplinary Theoretical and Mathematical Sciences (iTHEMS), RIKEN, Wako 351-0198, Japan}

\author{Boseong Kim}
\affiliation{IBM Quantum, IBM Japan, Tokyo 103-8510, Japan}

\author{Ryo Wakizaka}
\affiliation{IBM Quantum, IBM Japan, Tokyo 103-8510, Japan}

\author{Ikko Hamamura}
\affiliation{NVIDIA Corporation, ATT EAST 12F, 2-11-7 Akasaka, Minato-ku, Tokyo 107-0052, Japan}

\author{Mitsuhisa Sato}
\affiliation{RIKEN Center for Computational Science (R-CCS), Kobe 650-0047, Japan}

\author{Seiji Yunoki}
\affiliation{RIKEN Center for Computational Science (R-CCS), Kobe 650-0047, Japan}
\affiliation{Computational Condensed Matter Physics Laboratory, RIKEN Pioneering Research Institute (PRI), Wako 351-0198, Japan}
\affiliation{RIKEN Center for Quantum Computing (RQC), Wako 351-0198, Japan}
\affiliation{RIKEN Center for Emergent Matter Science (CEMS), Wako 351-0198, Japan}

\author{Mario Motta}\thanks{Corresponding authors.}
\affiliation{IBM Quantum, IBM Research, Thomas J. Watson Research Center, Cambridge, MA 02142, USA}
\email{mario.motta@ibm.com}

\author{Hiroshi Horii}
\affiliation{IBM Quantum, IBM Japan, Tokyo 103-8510, Japan}

\author{Yuri Kobayashi}
\affiliation{IBM Quantum, IBM Japan, Tokyo 103-8510, Japan}

\date{\today}

\begin{abstract}
For strongly correlated systems such as iron--sulfur clusters, the accuracy of
sample-based quantum diagonalization (SQD) has remained limited. The accuracy of these
calculations is affected by the mean-field reference state, which determines both the sampling
ansatz and the target of the configuration-recovery
step. We extend the \texttt{qcsc-prefect} quantum-centric supercomputing workflow---and its
reference-preparation stage for SQD---to support a broken-symmetry unrestricted
Hartree--Fock (BS-UHF) starting point, i.e., an unrestricted Hartree--Fock (UHF) solution with
broken spin symmetry obtained from atom-localized antiferromagnetic initial guesses. On the
$[\mathrm{Fe}_4\mathrm{S}_4(\mathrm{SCH}_3)_4]^{2-}$ cubane (54 electrons, 36 orbitals, 72 data qubits; 88 total qubits),
this single change lowers the mean-field reference by $\approx$534~mHa relative to the Restricted Hartree–Fock (RHF) solution ($-327.0809$ vs.\ $-326.5474$~Ha) and,
combined with deep multi-step configuration recovery on ROQUO graphics processing unit (GPU) nodes
at subspace dimension $d=3\times10^{10}$, reaches
$E=-327.235076$~Ha, our deepest directly computed point ($4.3$~mHa above the
finite-$D=4000$ density-matrix renormalization group (DMRG) value, $-327.2393952$~Ha) and $\sim$512--517~mHa below a same-active-space
RHF-reference SQD calculation reported earlier on the identical cluster. Separately, extrapolating across subspace dimension to the $d\to\infty$
limit via a Gaussian-in-$\log d$
fitting ansatz (commonly used for bond-dimension or subspace-size extrapolations) gives an
asymptotic estimate
$E_\infty=-327.244468\pm1.05$~mHa, $0.33$~mHa above our own
discarded-weight-extrapolated DMRG estimate ($E_{D\to\infty}=-327.244797\pm0.46$~mHa), a difference
smaller than their combined $1\sigma$ uncertainty ($\approx1.15$~mHa). We characterize the
determinant-subspace composition driving this convergence, show a matching validation on
$[\mathrm{Fe}_2\mathrm{S}_2(\mathrm{SCH}_3)_4]^{2-}$, and report matched multi-high-performance-computing
(HPC) execution of the \texttt{qcsc-prefect} workflow on Fugaku, Miyabi-G, and ROQUO. The central conclusion
is that reference-state preparation -- not subspace size alone -- is a key driver of SQD energy
accuracy in the presence of strong electronic correlation.
\end{abstract}

\maketitle

\section{Introduction}

Sample-based quantum diagonalization (SQD) combines a quantum processor with classical high-performance computing (HPC) to estimate ground- and excited-state properties for active spaces of size beyond the reach of exact diagonalization~\cite{robledomoreno2025}. In the quantum-centric supercomputing (QCSC) paradigm, quantum sampling and classical numerical linear algebra are orchestrated as a single workflow, so that cost/accuracy trade-offs are determined by end-to-end scheduling and resource allocation rather than by either platform in isolation~\cite{robledomoreno2025,shirakawa2025,kanazawa2025,qcsc_2506_20658,qcsc_methylene_jctc_2025,prresearch_75pv_hbrx,prresearch_dmn4_snfx}.

This systems view builds on decades of HPC workflow engineering, where throughput depends on queueing policies, co-scheduling, and concurrent execution across heterogeneous resources~\cite{hpc_sched_2506_10052,future_generation_computer_systems_2024_03583,hpc_sched_2412_18519,acm_3731599_3767551}. Incorporating quantum hardware adds operational constraints (hardware availability, calibration/noise drift, and cloud access patterns) that must be handled alongside the classical workload~\cite{kanazawa2025,ieee_11018264}.

In SQD, one or more quantum circuits are executed on quantum hardware to generate bitstring samples. Those samples are then repaired toward the correct particle number by a self-consistent configuration-recovery procedure, and the Hamiltonian projected into the recovered determinant subspace is diagonalized on classical nodes (typically using a Davidson-type solver)~\cite{robledomoreno2025,kanazawa2025}. SQD is commonly paired with locality- and hardware-aware ansatz constructions, including the local unitary cluster Jastrow (LUCJ) family~\cite{lucj_2511_22476,rsc_51_5_1659,motta_jctc_2019_ucj}. Such locality assumptions reduce two-qubit routing overhead on low-connectivity devices and can decrease the number of required SWAP operations~\cite{rsc_sc_14_40_11213,pmc11866757}. Practical workflows often initialize variational parameters from a classical coupled-cluster singles and doubles (CCSD) calculation~\cite{ccsd_rmp_2007,shirakawa2025}, an initialization we adopt here using the improved LUCJ parameter-fitting procedure of Ref.~\cite{lucj_2511_22476}.

Robledo-Moreno~et al. demonstrated SQD on N$_2$ and on $[\mathrm{2Fe\text{-}2S}]$ and $[\mathrm{4Fe\text{-}4S}]$ iron--sulfur clusters by pairing a quantum processor with Fugaku~\cite{robledomoreno2025}. Shirakawa~et al. later introduced a closed-loop strategy---a differential-evolution outer optimization wrapped around the sampling/recovery/diagonalization inner loop---and scaled it to the full supercomputer, engaging up to 152{,}064 Fugaku nodes~\cite{shirakawa2025}. Kanazawa~et al. added an observability layer for such workflows and reported repeated SQD executions on $[\mathrm{Fe}_4\mathrm{S}_4(\mathrm{SCH}_3)_4]^{2-}$ starting from a restricted Hartree--Fock (RHF) reference~\cite{kanazawa2025,arxiv_2602_01165}.

Iron--sulfur clusters are central to bioinorganic chemistry and have long served as stringent benchmarks for strongly correlated electronic-structure methods~\cite{Mende2021,sharma2014nchem}. Their low-energy manifolds often exhibit pronounced multireference character, making calculated results unusually sensitive to the mean-field reference used to initialize correlated treatments~\cite{Li2019}. For the tetrameric $[4\mathrm{Fe\text{-}4S}]$ cubane, Noodleman's broken-symmetry (BS) construction provides a useful approximate mean-field picture of locally high-spin iron centers coupled antiferromagnetically to a global singlet~\cite{noodleman1981}. The BS determinant is not spin adapted and does not replace a rigorous configuration-state-function or multireference description, but it can capture local moments and magnetic correlations absent from a single RHF closed-shell determinant. If the target quantity is spin-state gaps, spectra, or other properties that depend on a definite total spin, a spin-eigenstate (spin-adapted) treatment is required instead; the present work targets ground-state energies, for which a broken-symmetry reference is a useful and computationally accessible starting point.

\begin{figure}[t]
\centering
\includegraphics[width=\columnwidth]{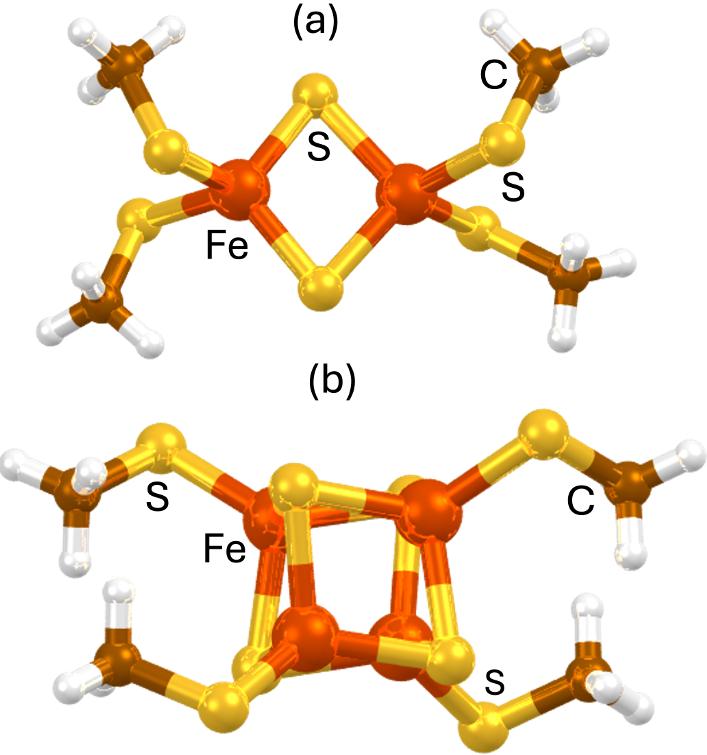}
\caption{Geometries of the iron--sulfur clusters studied in this work: (a) the dimeric
$[\mathrm{2Fe\text{-}2S}]$ cluster and (b) the tetrameric $[\mathrm{4Fe\text{-}4S}]$ cluster. Iron is
shown in orange, sulfur in yellow, carbon in brown, and hydrogen in white.}
\label{fig:fes_clusters}
\end{figure}

The electronic Hamiltonian has a symmetry group associated with conservation of the total spin (total magnetic momentum); any exact eigenstate of $\hat H$ also commutes with, and so can be labeled by an eigenvalue of, the total-spin operator $\hat S^2=\hat{\vec S}\cdot\hat{\vec S}$, with eigenvalue $S(S+1)$ for an exact total-spin quantum number $S$. For a general trial wavefunction -- in particular a single Slater determinant that is not spin-adapted, such as the unrestricted and broken-symmetry (BS) determinants used throughout this work -- $|\Psi\rangle$ is not an eigenstate of $\hat S^2$, so $\langle\hat S^2\rangle\equiv\langle\Psi|\hat S^2|\Psi\rangle$ is the expectation value of this operator rather than a sharp quantum number: it need not equal $S(S+1)$ for any integer or half-integer $S$, and a mixed value (e.g., $\langle\hat S^2\rangle=4.893$, between the $S=1$ value of $2$ and the $S=2$ value of $6$) signals that $|\Psi\rangle$ is a superposition of pure-spin components rather than a single multiplet -- this is what we mean by spin contamination throughout. $\langle\hat S^2\rangle$ is therefore a single global scalar describing the reference determinant as a whole; it is distinct from, and does not by itself report, the local spin populations on individual metal centers or the pairwise spin-correlation functions between them -- quantities we also discuss qualitatively (e.g., antiferromagnetic coupling between specific Fe centers) but do not separately compute here. In $[4\mathrm{Fe\text{-}4S}]$ and related clusters, this symmetry is itself one of the sources of strong correlation: multiple locally high-spin Fe centers couple antiferromagnetically to form a total spin-zero (or other low-spin) state, and representing that coupled state mathematically requires combining many configurations -- precisely the regime in which SQD's determinant-subspace diagonalization, and near-exact methods such as DMRG, are needed in place of a single mean-field determinant. In a mean-field or correlated (e.g., SQD) calculation, this spin symmetry can be either broken or preserved by the reference used to seed the calculation, and it is not obvious a priori which choice is more effective in a generic strongly correlated system -- this is the L\"owdin symmetry dilemma: the choice between symmetry-adapted and symmetry-broken references reflects a genuine dilemma, rather than a simple distinction between a correct and an incorrect mean-field state. For the nitrogenase FeMo cofactor, spin-unrestricted orbitals absorb much of the spin--charge correlation into the basis itself, markedly improving DMRG energies at fixed bond dimension~\cite{zhai2026femoco}. Motivated by this, we explore the use of broken-symmetry determinants to set up SQD calculations. Here we assess whether a BS reference can improve SQD performance by enhancing sampling and determinant-subspace recovery, thereby reducing the residual discrepancy to a near-exact DMRG reference relative to an RHF-initiated workflow. Figure~\ref{fig:fes_clusters} shows the two iron--sulfur motifs studied here.

Unrestricting the reference carries the cost usually raised against it: the determinant is no longer an eigenstate of $\hat S^2$ but a mixed-spin weighted average over pure multiplets~\cite{noodleman1981,yamaguchi1988}, and this spin contamination is the standard objection to unrestricted Hartree--Fock (UHF) and broken-symmetry DFT alike~\cite{neese2009}. We nonetheless adopt it, for two reasons. First, spin-up and spin-down electrons do not in general experience the same effective potential, least of all when their numbers differ; forcing them into common spatial orbitals is the artificial constraint, and relaxing it lets the mean field develop the local moments iron--sulfur clusters carry~\cite{sharma2014nchem,zhai2026femoco}. Second, the systems this method is built to serve are driven by redox events: electron transfer adds or removes an electron, so the relevant states are intrinsically open-shell with unequal $\alpha$ and $\beta$ populations, and their energetics are built from paired oxidized and reduced calculations~\cite{blumberger2015}. The resulting spin contamination is the expected price of a broken-symmetry reference.

Accordingly, we ask what fraction of the remaining gap to the near-exact density-matrix renormalization group
(DMRG) reference is attributable to finite sampling and determinant-subspace truncation, and what fraction can be mitigated by a
reference that explicitly encodes the expected local-moment and antiferromagnetic-correlation pattern.

We extend \texttt{qcsc-prefect}\footnote{\url{https://github.com/kunalroy332/qcsc-prefect}}, the
workflow underlying Refs.~\cite{robledomoreno2025,shirakawa2025,kanazawa2025}, by updating its SQD
reference-preparation stage to support a broken-symmetry unrestricted Hartree--Fock (BS-UHF)
starting point. We then quantify the resulting improvement in SQD energy accuracy, measured as the
absolute deviation of the SQD energy from the DMRG reference energy, on $[4\mathrm{Fe\text{-}4S}]$
at fixed sampled-subspace size.
Specifically, this single change---not a larger sampled subspace---closes the majority of the
remaining error on $[4\mathrm{Fe\text{-}4S}]$, reaching within 4.3~mHa of the finite-$D=4000$ DMRG value at the same
72-qubit active space size used in Ref.~\cite{kanazawa2025}.

\begin{figure}[t]
\centering
\includegraphics[width=\columnwidth]{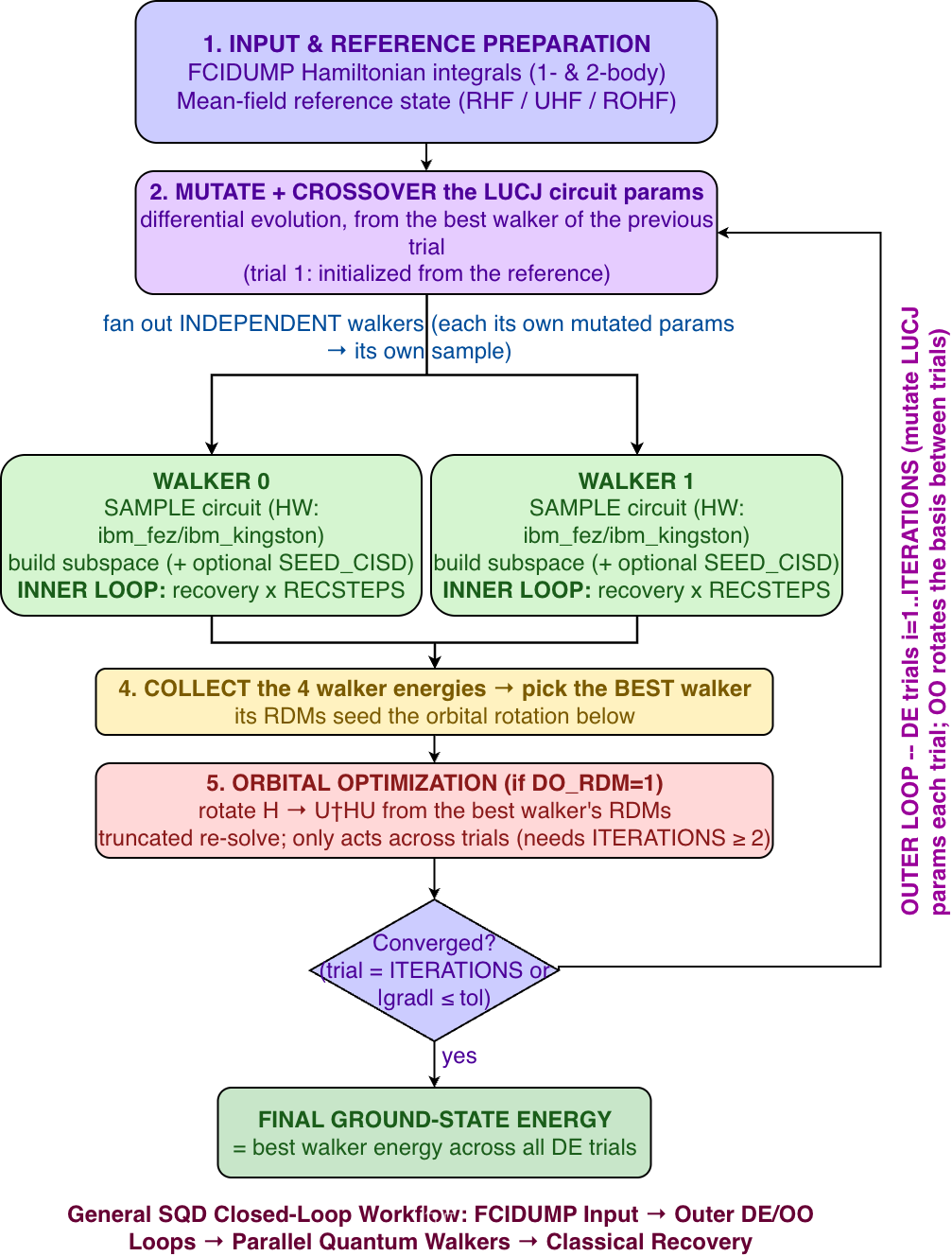}
\caption{The \texttt{qcsc-prefect} workflow used to run SQD calculations for
$[4\mathrm{Fe\text{-}4S}]$. (1) Reference-state preparation
constructs an atom-localized Noodleman BS-UHF guess (Fe1/Fe2 up, Fe3/Fe4 down) using an
antiferromagnetic (AF) Fe-site grouping and an initial spin-polarization pattern, and by allowing
spin-symmetry breaking on the bridging sulfurs in unrestricted Hartree--Fock (UHF). (2) An outer
differential-evolution (DE) loop mutates and recombines the LUCJ ansatz's variational parameters,
initialized from the reference on trial~1 and from the previous trial's best walker thereafter.
(3) Each trial fans out to independent walkers, each sampling its own mutated circuit on real
hardware, optionally seeding the recovered subspace with classical configuration-interaction (CI)
strings (see the seed configuration interaction with singles and doubles (CISD) ablation in
Sec.~\ref{sec:results}), and running its own inner configuration-recovery loop for a specified
number of recovery steps. (4) The best-energy walker's one- and two-particle reduced density
matrices (1-/2-RDMs) are collected. (5) If orbital optimization is enabled, those RDMs rotate the
active-space Hamiltonian $\hat H\to U^\dagger \hat H U$ for a truncated re-solve before the next DE
trial. The loop repeats until the trial budget is exhausted or the parameter gradient falls below
tolerance; the final reported energy is the best walker across every DE trial.}
\label{fig:pipeline}
\end{figure}

\section{The \texttt{qcsc-prefect} Workflow}

Figure~\ref{fig:pipeline} sketches the \texttt{qcsc-prefect} workflow as presented in
Ref.~\cite{kanazawa2025}; we use it here as a schematic overview. A reference Hartree--Fock state $|\Phi_0\rangle$ determines (i) the LUCJ ansatz
circuit's initial occupation and its orbital-rotation/Jastrow parameters, and (ii) the target
particle number and an initial-occupancy prior, supplied by the user and refined by configuration
recovery, used to repair noisy
bitstrings toward physical determinants; a low-diversity or otherwise poor prior degrades this
repair step. Recovered determinants populate a subspace
$\mathcal{S}_i$ at recovery step $i$; the Hamiltonian $\hat H$ projected onto $\mathcal S_i$ is
diagonalized via Davidson on a distributed classical solver (\texttt{sbd}~\footnote{\url{https://github.com/r-ccs-cms/sbd}}), and a carryover set
of high-weight determinants from step $i$ seeds step $i+1$ together with fresh samples, so that
useful configurations accumulate rather than being resampled from scratch every step. This inner
recovery loop runs independently inside each of several parallel walkers, each carrying its own
differential-evolution-mutated copy of the ansatz parameters (initialized from $|\Phi_0\rangle$ on
the first outer trial); the best-energy walker's recovered 1-/2-RDMs are what an optional outer
orbital-rotation step (\S\ref{sec:outlook}) would act on before the next outer trial, and the
pipeline's final reported energy is always the best walker seen across every outer trial.

Before this work, \texttt{qcsc-prefect} supported closed-shell RHF and restricted open-shell
Hartree--Fock (ROHF) references natively and provided a
nominal ``UHF'' mode that, for the systems considered here, often converged to the same stationary
solution as RHF. For readers less familiar with Hartree--Fock theory: an HF calculation minimizes
an energy functional over single Slater determinants, optimizing a set of molecular orbitals
subject to the constraint chosen for the reference -- identical (restricted) spatial orbitals for
$\alpha$- and $\beta$-spin electrons in RHF/ROHF, or independent (unrestricted) spatial orbitals for
each spin in UHF. This minimization is a nonlinear, non-convex self-consistent-field (SCF)
procedure, and the SCF iterations for a nominally unrestricted calculation can converge onto a
restricted solution rather than a genuinely spin-polarized one, particularly when the initial guess
carries no spin-symmetry-breaking bias. This behavior follows from a standard property of the Hartree--Fock energy
landscape: a closed-shell RHF solution is also a stationary point of the more general unrestricted
Hartree--Fock (UHF) functional. Consequently, initializing UHF with identical $\alpha$- and
$\beta$-spin densities provides no explicit spin-symmetry-breaking bias, and the SCF iterations may
remain at, or return to, the spin-pure RHF solution, with $\langle \hat S^2\rangle = 0$ and an energy
indistinguishable from RHF. This convergence to the RHF solution can be avoided by an appropriate
choice of initial guess or, as described next, by explicit stability analysis.

Stability analysis requires a more precise distinction. Internal RHF stability analysis explores
orbital rotations that remain within the restricted manifold and therefore cannot, by itself,
generate a broken-spin UHF determinant. External RHF stability analysis, in contrast, probes
symmetry-breaking directions that leave the RHF manifold (including RHF-to-UHF instabilities), and
following an unstable external mode can provide a fully unrestricted starting point. However,
for Fe--S cubanes this local, Hessian-based procedure does not guarantee convergence to a
chemically relevant antiferromagnetic minimum among many near-degenerate broken-symmetry basins. In
our calculations on $[4\mathrm{Fe\text{-}4S}]$, the spin-symmetric initialization remained in the
spin-pure basin ($E=-326.547$~Ha), and reproducibly reaching the targeted antiferromagnetic state
required an explicit atom-localized (Noodleman) spin-polarized initial density. Thus, the general
problem is not that external stability analysis is fundamentally unable to generate a UHF
state, but that controlling UHF convergence -- steering it away from the spin-pure solution that is
always accessible when $N_\alpha=N_\beta$, and toward the specific, chemically relevant
broken-symmetry basin -- requires a robust, chemically informed mechanism, which we provide within
\texttt{qcsc-prefect} so that users can run the best possible BS-SQD calculations.

We address this with a sequence of workflow-level changes spanning reference-state preparation
and the configuration-recovery loop: computation of unrestricted one- and two-body
integrals and unrestricted coupled-cluster singles and doubles (UCCSD) natural-orbital occupancies; an
antiferromagnetic promotion guess for Fe--S cubanes; an atom-localized (Noodleman) initial density
keyed to which magnetic
center each active orbital belongs to; fractional-polarization and free-bridging-ligand knobs
(\S\ref{sec:refstate}); internal and external stability analysis to ensure that each converged
determinant is locally stable, together with multiple chemically informed atom-localized initial
densities to locate and compare distinct near-degenerate broken-symmetry minima; this emphasis on reliably reaching
and comparing low-lying broken-symmetry states is consistent with the dense, multi-state
low-energy spectra for $[2\mathrm{Fe\text{-}2S}]$ and $[4\mathrm{Fe\text{-}4S}]$ clusters revealed by
DMRG calculations~\cite{sharma2014nchem}. In the configuration-recovery loop itself, we implemented a
carryover-forwarding fix, added checkpoint/restart capability to support extended executions, and
introduced per-excitation-rank (singles/doubles/triples/quadruples-plus) accounting of the recovered
subspace used in \S\ref{sec:composition}.

\section{Reference-State Preparation: The Broken-Symmetry Ansatz}
\label{sec:refstate}

The reference state is a broken-symmetry unrestricted Hartree--Fock (BS-UHF) solution obtained
using a standard fragment-based initialization and stability-following procedure: the active-space
orbitals are partitioned into disjoint fragments (one per metal center, one for the bridging
sulfides, one per spectator-ligand group), each fragment is assigned a spin role, and the resulting
spin-localized density seeds a UHF calculation using the one- and two-electron integrals of the
target \texttt{FCIDUMP}. The converged solution is then refined to a stable local minimum by
following both internal (spin-conserving) and external (spin-symmetry-breaking) instabilities.
Appendix~\ref{app:bsuhf} gives the construction explicitly, including the fractionally polarized
seed density of Eq.~\eqref{eq:afguess}, the stability-following recursions, and the convergence
path on $[4\mathrm{Fe\text{-}4S}]$, for completeness and reproducibility, as these procedures
follow standard practice for obtaining UHF solutions and are not themselves the focus of the
present work.

Before turning to the SQD results, two aspects of the BS-UHF reference require discussion, since
both set the baseline for every number reported in \S\ref{sec:results} and are not fixed by the
standard construction. First, for a four-metal cluster, the broken-symmetry ansatz admits several
distinct sublattice pairings that are nearly degenerate; selecting the physically relevant pairing
ultimately requires a correlated calculation rather than comparison of the mean-field energies
alone (\S\ref{sec:sweep}). Second, the reference is substantially spin-contaminated (more so than a
softer, less polarized seed would be), and it is therefore important to address this point
explicitly before using the state to seed a quantum sampling procedure.

The external half of the stability search is what makes the deeper basin reachable. We therefore ran
two otherwise identical calculations on $[4\mathrm{Fe\text{-}4S}]$ (same atom-localized initial guess,
same FCIDUMP, same Newton solver), differing only in whether external instabilities are followed.
Following internal (spin-conserving) instabilities alone, the guess relaxes to a local minimum
$219$~mHa below RHF ($E=-326.767$~Ha, $\langle\hat S^2\rangle\approx7.6$) where no internal
instability remains. When external (spin-symmetry-breaking) instabilities are followed as well, and
the unstable-mode orbitals are passed directly to the Newton solver, the procedure reaches
$E=-326.801$~Ha, i.e., $34$~mHa below the internal-only minimum. Thus the $34$~mHa separation is
attributable solely to crossing an external-instability barrier, not to any difficulty in SCF
convergence once past it.

\begin{figure}[t]
\centering
\includegraphics[width=0.82\columnwidth]{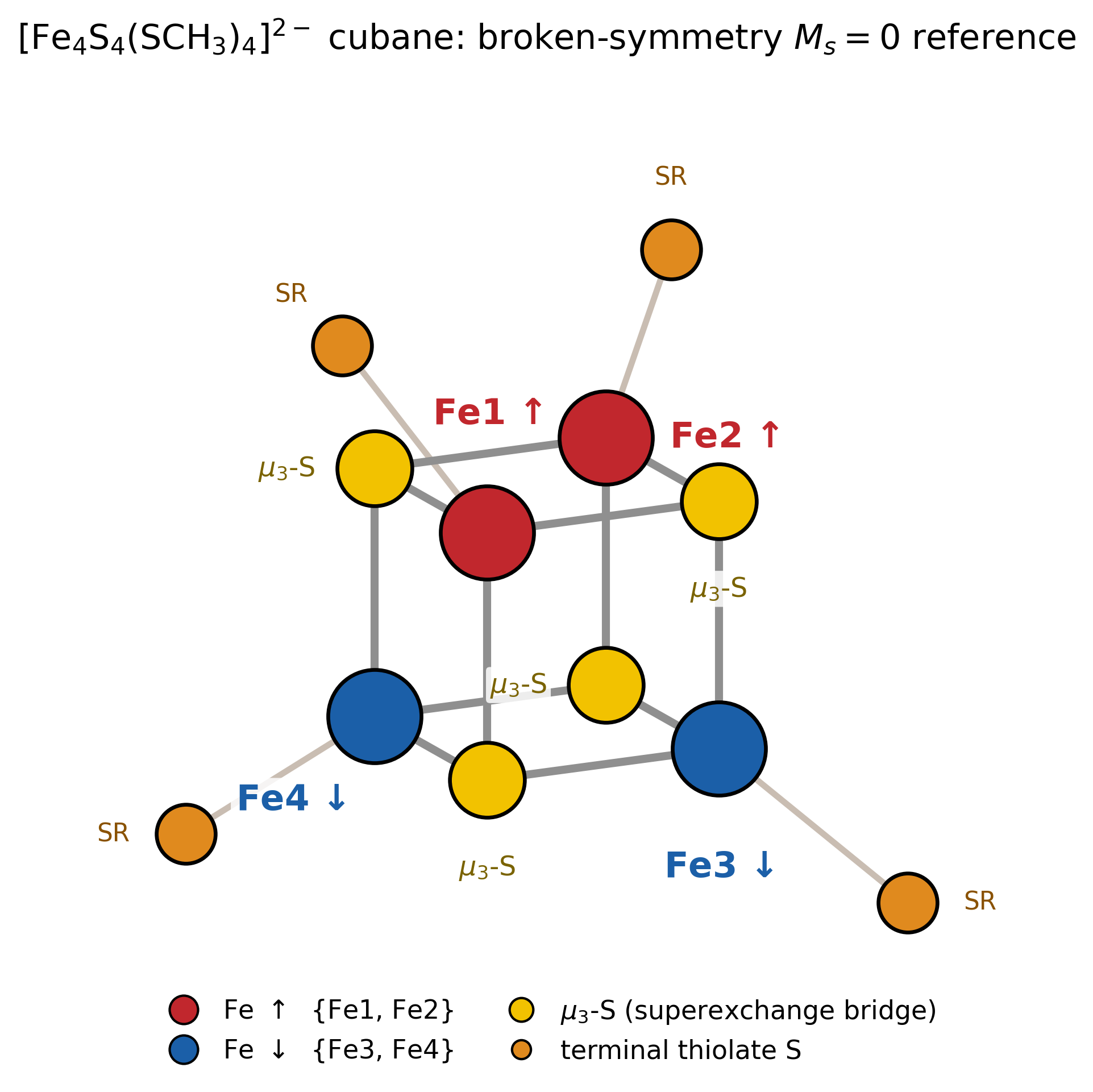}
\caption{The $[4\mathrm{Fe\text{-}4S}]$ cubane with the actual broken-symmetry pairing used to
generate every sampled pool and SQD result in this paper: Fe1/Fe2 spin-up (red), Fe3/Fe4 spin-down
(blue), the four $\mu_3$-bridging sulfides that mediate the antiferromagnetic superexchange
(yellow), and the four terminal thiolate ligands (orange). This is pairing (1,2) of
Table~\ref{tab:pairing-sweep} -- the correlated ground pairing identified in \S\ref{sec:sweep} --
not an arbitrary illustrative choice.}
\label{fig:cubane}
\end{figure}

\begin{figure}[t]
\centering
\includegraphics[width=\columnwidth]{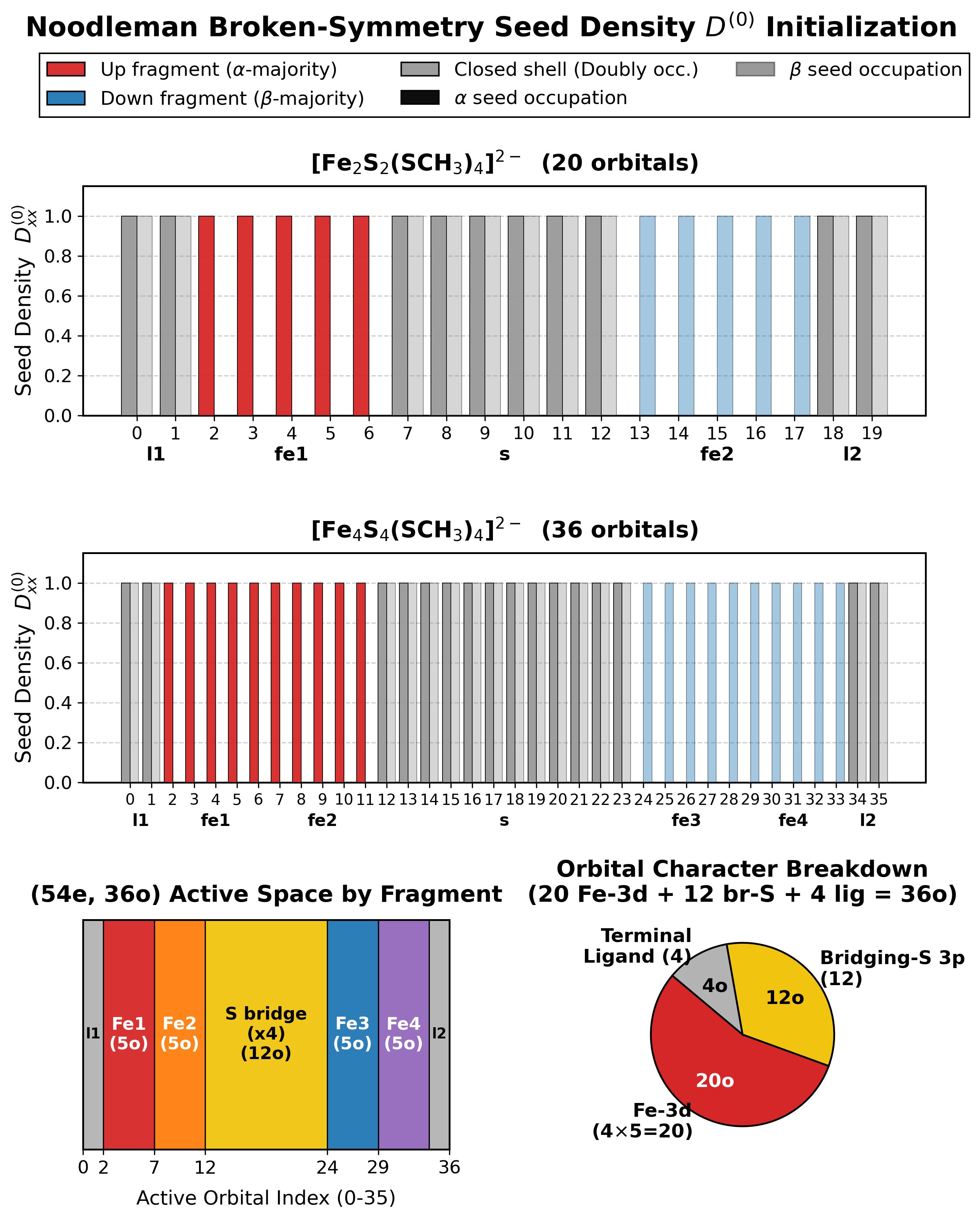}
\caption{Top two rows: the seed density $D^{(0)}$ of Eq.~\eqref{eq:afguess} at full polarization
($p=1$), concretely for both clusters studied: $[2\mathrm{Fe\text{-}2S}]$ (20 orbitals, one Fe
fragment per sublattice) and $[4\mathrm{Fe\text{-}4S}]$ (36 orbitals, two Fe fragments per
sublattice). Up fragments (red) start fully $\alpha$-occupied and $\beta$-empty; down fragments
(blue) the reverse; everything else (ligand and bridging-sulfide orbitals, gray) starts
closed-shell -- the literal input handed to $\mathrm{UHF}(D^{(0)})$ before any self-consistent-field
(SCF) iteration or stability-following. Bottom row: how the $[4\mathrm{Fe\text{-}4S}]$ active space
(54 electrons, 36
orbitals) decomposes by fragment and orbital character -- four Fe $3d$ manifolds (5 orbitals each,
20 total), one 12-orbital bridging-sulfide block shared across all four $\mu_3$-S, and 4
spectator-ligand orbitals -- the same fragment map Fig.~\ref{fig:cubane} draws structurally.}
\label{fig:localization}
\end{figure}

\subsection{The pairing sweep and a deeper reference}
\label{sec:sweep}

The internal+external stability-following procedure above was subsequently repeated on a second,
independently constructed
$[4\mathrm{Fe\text{-}4S}]$ active-space \texttt{FCIDUMP}, built from a more strongly localized
active space than the one used above (with the core energy set to zero). Unless otherwise noted, every result from
\S\ref{sec:results} onward is built on this updated reference, not the $E=-326.801$~Ha value above.
Because the Noodleman construction admits three distinct AF sublattice pairings for four Fe
centers, namely $(1,2)=\{\mathrm{Fe1,Fe2}\}$ up, $(1,3)=\{\mathrm{Fe1,Fe3}\}$ up, and
$(1,4)=\{\mathrm{Fe1,Fe4}\}$ up, we examined all three. Each pairing
was run under two seed densities: full polarization
($p=1$ in Eq.~\eqref{eq:afguess}, i.e., one electron per spin channel fully confined to its
sublattice) and a softened guess ($p=0.6$ with the bridging-sulfide fragment left unpolarized) that
had helped reach a comparable basin on the earlier FCIDUMP. Each case was then carried through
unrestricted coupled-cluster singles and
doubles (UCCSD) and its perturbative-triples correction, UCCSD(T) (Table~\ref{tab:pairing-sweep};
Fig.~\ref{fig:cubane} shows pairing (1,2) structurally).

\begin{table}[t]
\caption{UHF/UCCSD/UCCSD(T) energies (Ha) for all three Noodleman broken-symmetry sublattice
pairings on the
localized \texttt{FCIDUMP}, at full polarization and for the softened initial guess (abbreviated
SP: $p=0.6$ with the bridging-sulfide fragment left free/unpolarized). Pairing $(i,j)$ denotes the
spin-up sublattice $\{\mathrm{Fe}i,\mathrm{Fe}j\}$, with the complementary pair spin-down.
Fragment-orbital
ranges are Fe1~2--6, Fe2~7--11, Fe3~24--28, and Fe4~29--33, as indicated in Fig.~\ref{fig:localization}.}
\label{tab:pairing-sweep}
\footnotesize
\makebox[\linewidth][l]{%
\begin{tabular}{lccccc}
\hline\hline
Pairing & guess & UHF & $\langle\hat S^2\rangle$ & UCCSD & UCCSD(T)\\
\hline
(1,2) & full pol. & $-327.0809$ & $8.877$ & $-327.2096$ & $-327.2219$\\
(1,3) & full pol. & $-327.0791$ & $8.874$ & $-327.2093$ & $-327.2219$\\
(1,4) & full pol. & $-327.0889$ & $8.744$ & $-327.1980$ & $-327.2120$\\
(1,2) & SP & $-327.0204$ & $8.403$ & $-327.1751$ & $-327.2107$\\
(1,3) & SP & $-327.0202$ & $8.402$ & $-327.1750$ & $-327.2107$\\
(1,4) & SP & $-327.0889$ & $8.744$ & $-327.1980$ & $-327.2120$\\
\hline\hline
\end{tabular}%
}
\end{table}

Two principal conclusions emerge from this sweep. First, the three nominal-singlet broken-symmetry
solutions are near-degenerate, and their ordering changes between the SCF and coupled-cluster
levels: the UHF energies rank $(1,4)$ lowest
($-327.0889$~Ha), but correlated UCCSD(T) ranks $(1,2)\equiv(1,3)$ lowest instead
($-327.2219$~Ha, with $(1,4)$ $\sim$10~mHa higher) -- $(1,2)$ and $(1,3)$ are energetically
indistinguishable at the reported precision and appear to represent symmetry-related AF
arrangements reached under different Fe labelings, and this correlated ordering, not the SCF-level
ordering, is what selects the ground pairing here. Because UCCSD(T) is not variational, a lower
UCCSD(T) energy is not guaranteed in general to indicate the more physically correct pairing; in
this case it is the more discriminating and higher-level criterion available to us, but this need
not hold for every system. Second,
the softened-polarization (SP) guess that helped reach a comparable basin
on the earlier non-localized FCIDUMP does not transfer here: it only moves $\langle\hat S^2\rangle$ from
$8.88$ to $8.40$ (still far from the $\sim6.26$ reference estimate)~\cite{sharma2014nchem} while costing
$\sim$60~mHa (UHF) and $\sim$11~mHa (UCCSD(T)), and stability-following from the softened
$(1,4)$ guess relaxes into the identical basin as full-polarization $(1,4)$. Full polarization
is therefore the right recipe on this \texttt{FCIDUMP}, and pairing $(1,2)$ (Fig.~\ref{fig:cubane})
is the reference used throughout \S\ref{sec:results}.

\begin{table}[t]
\centering
\caption{Comparison of two \texttt{FCIDUMP}s constructed for $[4\mathrm{Fe\text{-}4S}]$ that differ
in how strongly the active-space orbitals are localized: an earlier, non-localized active space
(``Old best'') versus a subsequently constructed, more strongly localized active space (``New
best''), used from \S\ref{sec:results} onward. Both sets of energies use the same core-energy
convention, $E_{\mathrm{core}}=0$, which
permits comparison without an additive core-energy offset. Because the two \texttt{FCIDUMP}s differ
in orbital localization, the reported differences $\Delta$ reflect both the resulting change in the
accessible mean-field basin and the subsequent correlation treatment, not a single isolated cause.
The ``New best'' column corresponds to the
full-polarization, pairing-$(1,2)$ broken-symmetry solution on the updated \texttt{FCIDUMP}.}
\label{tab:old-vs-new}
\begin{tabular}{lccc}
\hline\hline
Method & \shortstack{Old\\best} & \shortstack{New best\\(pairing $(1,2)$)} & $\Delta$\\
\hline
BS-UHF & $-326.801$ & $-327.081$ & $-280$~mHa\\
BS-UCCSD & $-327.143$ & $-327.209$ & $-66$~mHa\\
BS-UCCSD(T) & $\approx-327.18$ & $-327.222$ & $-42$~mHa\\
\hline\hline
\end{tabular}
\end{table}

Relative to the earlier, more naively prepared reference, the updated reference converges to a
mean-field solution of substantially lower energy and yields improved energies at each level of theory considered ("deeper" throughout this section refers to mean-field/correlated energy, not to any property of a quantum circuit). $\mathrm{UCCSD(T)} = -327.222$~Ha on this \texttt{FCIDUMP} is the target every SQD trajectory in
\S\ref{sec:results}--\ref{sec:hpc} is measured against. What is established here is that the
updated reference is reproducible, reaches a systematically lower-energy mean-field solution, and that the correlated ground-state
pairing is identified.

Notably, the deeper reference exhibits larger spin contamination; understanding the origin of this trend is important. A
broken-symmetry singlet is not a spin eigenstate: $\langle\hat S^2\rangle$ lies between the target
value ($0$) and that of the higher multiplicities it admixes. For this tetrameric
$[4\mathrm{Fe\text{-}4S}]$ cluster, Ref.~\cite{sharma2014nchem} reports an effective low-energy
estimate of $\langle\hat S^2\rangle\approx6.26$. Neither broken-symmetry reference reproduces this
estimate: the earlier nonlocalized reference gives approximately $7.6$, whereas the updated
reference gives approximately $8.87$.
Stronger localization produces more atomic-like Fe $3d$ orbitals and
therefore more fully developed local magnetic moments, approaching the high-spin Fe centers
expected from chemical intuition. This increased spin polarization raises the degree of spin
contamination in the broken-symmetry determinant. By contrast, a softer mean-field solution with
greater Fe--S covalency partially suppresses the local moments and therefore exhibits a smaller
$\langle\hat S^2\rangle$ value. The increase from $7.6$ to $8.87$ should therefore not necessarily
be viewed as an independent deficiency of the updated reference, but rather as a consequence of
the stronger local-moment character associated with the more localized orbital representation and
the lower-energy mean-field basin reached from it.

The BS-UHF determinant is used only as a starting point for
the subsequent correlated treatment: it provides the LUCJ
amplitudes and seeds the sampling procedure, while the Davidson diagonalization is carried out in
a fixed-$S_z$ subspace. Fixing $S_z$ does not by itself enforce a definite value of total spin
$S$, so the lower SQD energy alone does not establish that spin contamination has been resolved.
Any residual spin contamination in the final SQD state must be assessed directly,
for example through $\langle\hat S^2\rangle$, or controlled through an explicitly spin-adapted or
spin-projected treatment -- neither of which we undertake in this work. The more meaningful
energetic indicator is instead the approximately $42$~mHa
lowering of the UCCSD(T) energy on the updated, localized \texttt{FCIDUMP} relative to its value on
the earlier, non-localized one (Table~\ref{tab:old-vs-new}), which suggests that the
correlated treatment benefits from the improved
reference despite the greater spin contamination of the underlying mean-field determinant.

\section{Results}
\label{sec:results}

\subsection{$[\mathrm{Fe}_2\mathrm{S}_2(\mathrm{SCH}_3)_4]^{2-}$: validation}
\label{sec:fe2s2}

We begin our exploration of BS-SQD with $[2\mathrm{Fe\text{-}2S}]$, where the two Fe centers are
antiferromagnetically coupled: a smaller-scale system on which we apply the
reference-state preparation and recovery procedure of \S\ref{sec:refstate}, using 20 spatial orbitals, 40 qubits, and a CAS(30e,20o) active space.
Localizing the initial density onto one Fe
fragment per magnetic sublattice (Eq.~\eqref{eq:afguess}, Fig.~\ref{fig:localization}, top panel)
and following the external instabilities yields a BS-UHF solution with $E=-116.512692$~Ha
($\langle\hat S^2\rangle=4.893$) -- $377.7$~mHa below naive UHF, which, exactly as for
$[4\mathrm{Fe\text{-}4S}]$, collapses onto the closed-shell RHF solution
($E=-116.135035$~Ha, $\langle\hat S^2\rangle=0$) when seeded from the default energy-ordered guess
instead. UCCSD based on the BS-UHF reference reaches $-116.580470$~Ha. For comparison, the same
unrestricted coupled-cluster machinery applied instead to the closed-shell RHF reference gives
UCCSD $=-116.331276$~Ha and UCCSD(T) $=-116.37455$~Ha, while the previously reported near-exact
DMRG value is $-116.595$~Ha~\cite{sharma2014nchem}. Table~\ref{tab:fe2s2-refs} collects these
reference values for quick comparison.

\begin{table}[t]
\centering
\caption{$[2\mathrm{Fe\text{-}2S}]$ reference energies (Ha) and mean-field $\langle\hat S^2\rangle$
for the CAS(30e,20o) active space discussed in this section. ``UHF, naive'' is the nominal UHF
mode seeded from the default energy-ordered guess, which collapses onto the RHF solution (as for
$[4\mathrm{Fe\text{-}4S}]$); ``UHF, BS'' is the broken-symmetry solution used to seed SQD.
``UCCSD''/``UCCSD(T)'' use the RHF reference; ``BS-UCCSD'' uses the BS-UHF reference. BS-SQD is
this work's deepest directly computed recovery point (\S\ref{sec:fe2s2}); ``DMRG, $D=2400$ (this
work)'' is our own \texttt{block2} calculation at the largest sampled bond dimension
(Appendix~\ref{app:fe2s2dmrg}, Fig.~\ref{fig:fe2s2-dmrg}), lower than the extrapolated
$D\to\infty$ estimate quoted in the text since $E(D)$ decreases monotonically with $D$; a
BS-UCCSD(T) value and an independent, unextrapolated SQD result from an RHF-referenced literature
calculation are not available for direct inclusion here.}
\label{tab:fe2s2-refs}
\begin{tabular}{lcc}
\hline\hline
Method & $E$ (Ha) & $\langle\hat S^2\rangle$\\
\hline
RHF & $-116.135035$ & $0.000$\\
UHF, naive & $-116.135035$ & $0.000$\\
UHF, BS & $-116.512692$ & $4.893$\\
UCCSD & $-116.331276$ & ---\\
UCCSD(T) & $-116.37455$ & ---\\
BS-UCCSD & $-116.580470$ & ---\\
DMRG~\cite{sharma2014nchem} & $-116.595$ & $0.000$\\
BS-SQD (this work, deepest point) & $-116.603964$ & ---\\
DMRG, $D=2400$ (this work) & $-116.605417$ & $0.000$\\
\hline\hline
\end{tabular}
\end{table}

\begin{figure}[t]
\centering
\includegraphics[width=\columnwidth]{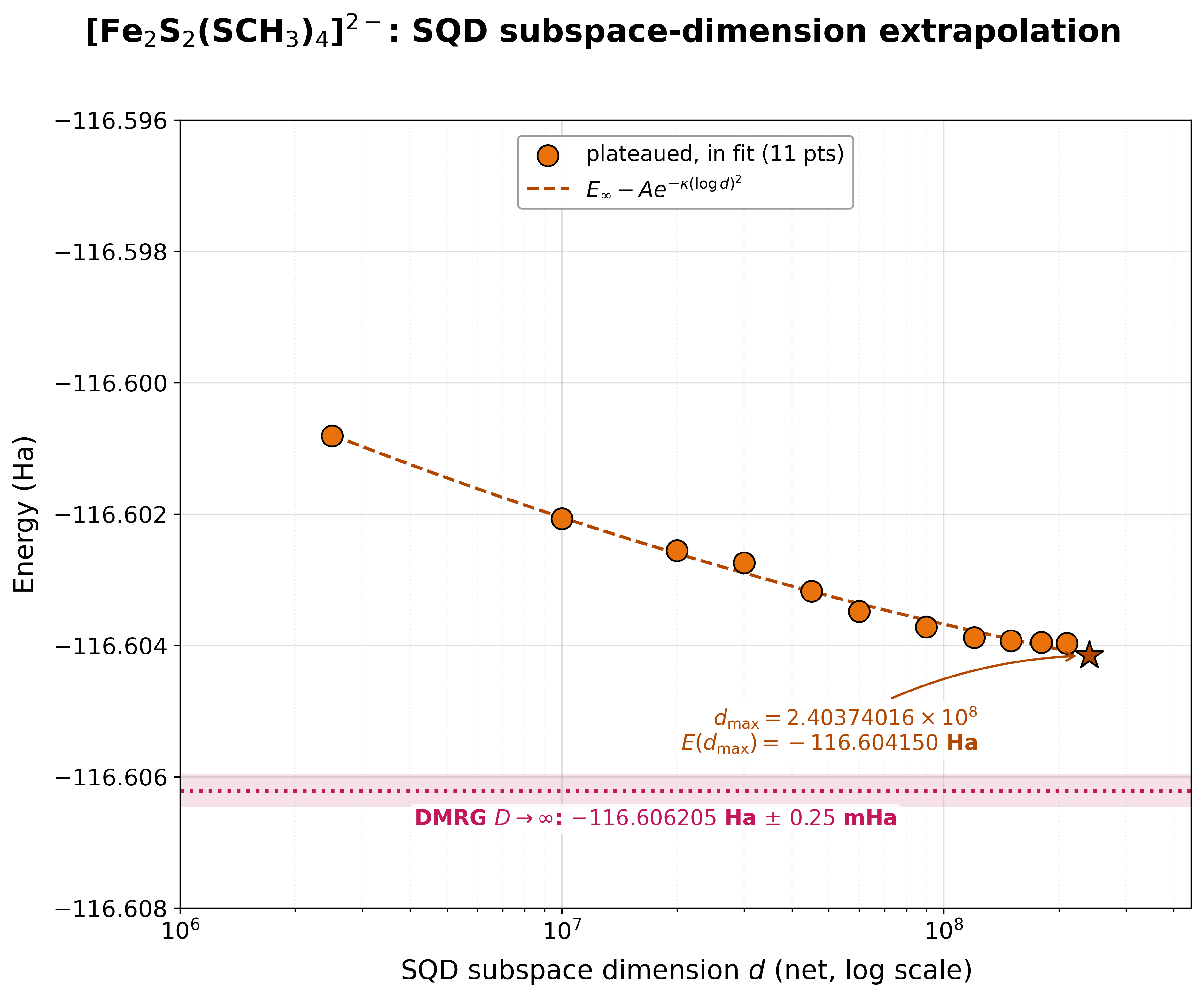}
\caption{$[2\mathrm{Fe\text{-}2S}]$: SQD energy vs.\ net subspace dimension $d$, spanning
$d=2.5\times10^{6}$ to $2.1\times10^{8}$; the points shown (orange circles) all enter the fit of
$E(d)=E_\infty - A\,e^{-\kappa(\log d)^2}$~\cite{zhai2026femoco} (dashed curve). The star and
annotation on the curve mark the fitted energy at the largest physically realizable determinant
subspace, $d_{\max}=2.40374016\times10^8=\binom{20}{15}^2$ ($E(d_{\max})=-116.604150$~Ha). Horizontal
line and shaded band, with the accompanying numerical label: the independent DMRG $D\to\infty$
extrapolation, shown here as a benchmark reference and detailed in Appendix~\ref{app:fe2s2dmrg},
Fig.~\ref{fig:fe2s2-dmrg} ($-116.606205$~Ha $\pm0.25$~mHa, $1\sigma$). The BS-UHF
($-116.5127$~Ha) and BS-UCCSD ($-116.5805$~Ha) references lie above the plotted window. This SQD
extrapolation and the DMRG cross-check of Fig.~\ref{fig:fe2s2-dmrg} are compared directly in the
text below.}
\label{fig:fe2s2}
\end{figure}

We sampled a one-layer LUCJ circuit built directly from these BS-UCCSD amplitudes using the
fermionic-circuit simulation package \texttt{ffsim}~\cite{ffsim2026,ffsim_software} (via
\texttt{ffsim.UCJOpSpinUnbalanced}) and executed it on real \texttt{ibm\_fez} hardware at two shot
counts, 100{,}000 and 500{,}000 (dynamical-decoupling and measurement-twirling error mitigation,
LUCJ-aware
transpilation), of which reset mitigation retained 25,945 and 130,964 shots, respectively. Of
those retained shots, only $0.59\%$ ($153$ shots) and $0.67\%$ ($877$ shots) already satisfy the
target particle number exactly on both spin sectors ($N_\alpha=N_\beta=15$) without any
correction; the per-spin Hilbert space has $\binom{20}{15}=15{,}504$ configurations, so requiring
the correct particle number on both spins simultaneously is a much smaller target than either
spin sector alone. Configuration recovery is what bridges the remaining, much larger gap. As shown
in Fig.~\ref{fig:fe2s2}, a single recovery pass at the resulting net subspace dimension
($1732\times1732=2{,}999{,}824$) already reaches $E=-116.556678$~Ha (100k shots, $23.79$~mHa above
BS-UCCSD) and $-116.554245$~Ha (500k shots, $26.23$~mHa above BS-UCCSD), both below naive UHF. The
500k-shot single pass is $2.43$~mHa higher than the 100k-shot one despite the larger pool; because
both are subsampled by occupancy frequency down to the same fixed subspace dimension for this
one-pass comparison, this is consistent with ordinary variation between two single stochastic
subsample draws rather than a systematic effect of shot count. Reusing the same 500k-shot pool
with zero additional quantum-hardware calls and pushing the
recovery to $d=2\times10^{7}$ (three recovery steps, three K-batches) reaches
$E=-116.599974$~Ha, i.e., $4.97$~mHa below the quoted DMRG reference~\cite{sharma2014nchem}.

To characterize the subspace-size dependence properly, we then ran configuration recovery over a
single, fixed set of samples from quantum hardware (no new hardware calls) at eleven independent subspace dimensions spanning
$2.5\times10^{6}$ to $2.1\times10^{8}$, with the step count per dimension
set from the observed tail gradient rather than fixed across the sweep. The final per-step energy
change is below $0.017$~mHa for every dimension in the sweep except the largest
($d=2.1\times10^{8}$, two completed steps, final increment $0.029$~mHa). The recovered energy
decreases monotonically across the evaluated dimensions, from
$E=-116.600807$~Ha at $d=2.5\times10^{6}$ to $E=-116.603964$~Ha at the largest computed point,
$d=209{,}989{,}081$. For this active space, the largest physically realizable determinant subspace is
$d_{\max}=\binom{20}{15}^2=15{,}504^2=240{,}374{,}016=2.40374016\times10^{8}$, so the largest computed
point corresponds to $87.4\%$ of $d_{\max}$. $d_{\max}$ itself was not directly evaluated; the fitted
curve gives $E(d_{\max})=-116.604150$~Ha (see below).

Fitting these points to a Gaussian-in-$\log d$
decay, $E(d) = E_\infty - A\,e^{-\kappa(\log d)^2}$ ($\log$ natural) -- the same functional form
Ref.~\cite{zhai2026femoco} uses for DMRG bond-dimension extrapolation on related Fe-S active
spaces, adopted here empirically for its log-of-resource decay shape rather than derived
specifically for SQD subspace dimension -- gives (residual $0.083$~mHa RMS)
a residual bootstrap
(4000 refits) $E_\infty = -116.606298$~Ha with a statistical
$\pm0.73$~mHa $1\sigma$ interval (Fig.~\ref{fig:fe2s2}, shaded band). The largest computed point
already reaches $87.4\%$ of $d_{\max}$, with $E_\infty$ lying only $2.15$~mHa beyond
$E(d_{\max})$, so this extrapolation is a small, well-supported correction on top of a nearly
complete direct calculation.

As an independent cross-check, we ran DMRG for this active space using the
\texttt{block2} package~\cite{block2_2023} with SU(2)-spin-adapted symmetry, sweeping bond
dimension $D\in\{200,400,800,1200,1600,2000,2400\}$ and extrapolating $E(D)$ linearly in the
discarded weight (DW) -- the standard proxy for residual truncation error -- over $D=800$--$2400$
(dropping $D=400,200$, whose DW lies outside the linear regime of the larger-$D$ points). This
gives $E_{D\to\infty}=-116.606205$~Ha $\pm0.25$~mHa
(Appendix~\ref{app:fe2s2dmrg}, Fig.~\ref{fig:fe2s2-dmrg}); the quoted uncertainty is a residual
bootstrap (4000 refits) on this linear fit's intercept, the same procedure used for the SQD
extrapolation above.

At the largest physically realizable subspace dimension, $E(d_{\max})=-116.604150$~Ha lies
$2.06$~mHa above the DMRG $D\to\infty$ benchmark. The SQD $d\to\infty$ extrapolated energy,
$E_\infty=-116.606298$~Ha, is itself only $0.09$~mHa below the DMRG extrapolation, well within
the combined $1\sigma$ uncertainty of the two methods ($\pm0.73$~mHa and $\pm0.25$~mHa
respectively).

As a qualitative precedent, Ref.~\cite{zhai2026femoco} reports a sub-mHa difference
($0.13$~mHa, at $D=2400$) between finite-$D$ unrestricted DMRG and a more accurate spin-adapted
result for a distinct (20o,30e) Fe(III)--Fe(III) active space. Because that calculation concerns a
different molecule, active space, and reference procedure, we cite it only to illustrate that
sub-mHa finite-$D$ residual errors can persist in related transition-metal calculations, not as
validation of the present energies or uncertainty estimate.

The clean end-to-end validation on $[2\mathrm{Fe\text{-}2S}]$ (a 20-orbital, 30-electron active space) motivated committing the larger, multi-HPC deep-recovery campaign of \S\ref{sec:results}--\ref{sec:hpc} to the substantially larger $[4\mathrm{Fe\text{-}4S}]$ active space (36 orbitals, 54 electrons), for which the near-exact DMRG cross-check of the kind used above is far more computationally demanding to converge tightly.

\subsection{$[\mathrm{Fe}_4\mathrm{S}_4(\mathrm{SCH}_3)_4]^{2-}$: tetrameric iron cluster}

\begin{figure*}[t]
\centering
\includegraphics[width=\textwidth]{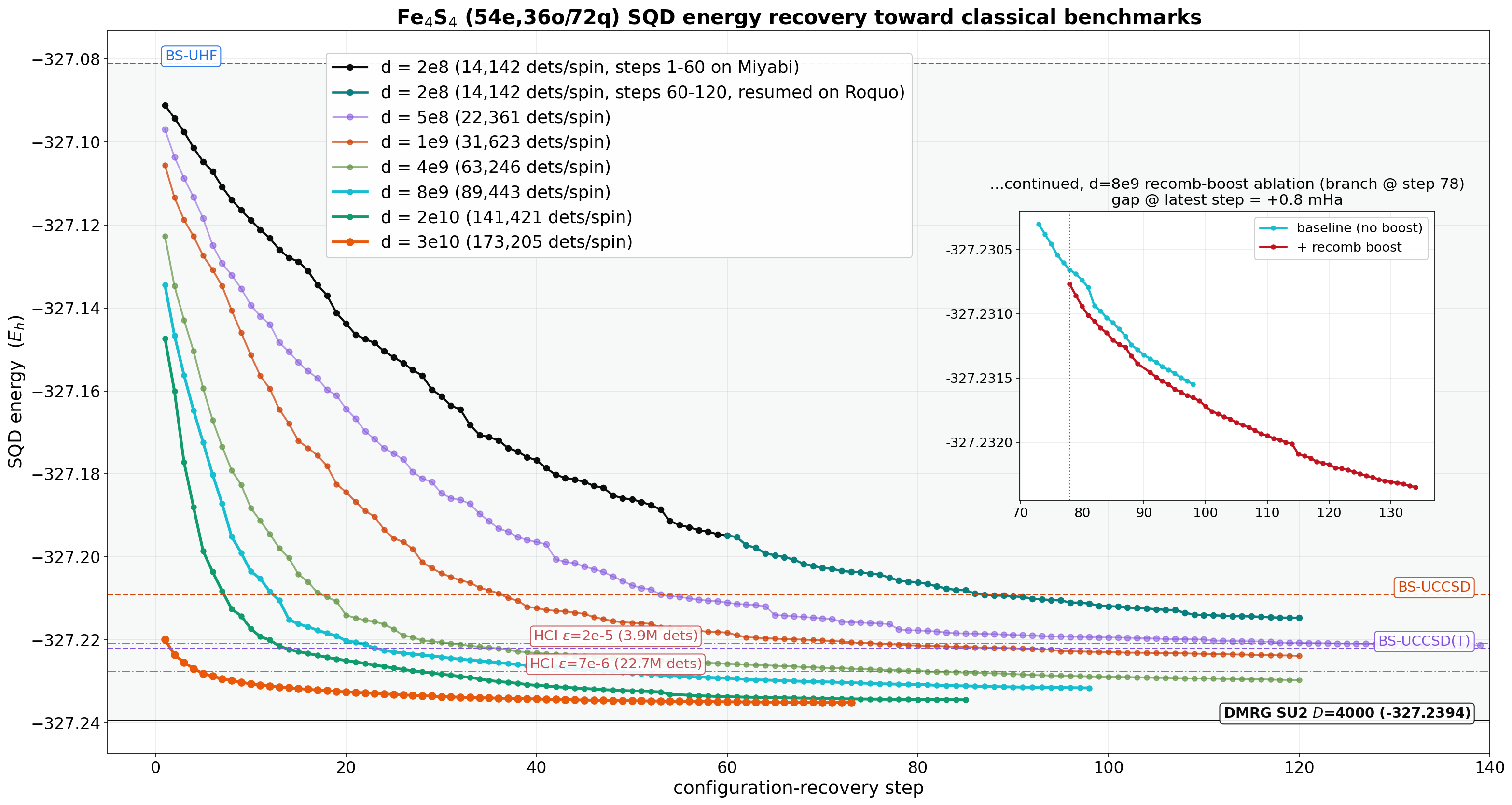}
\caption{$[4\mathrm{Fe\text{-}4S}]$ SQD energy vs.\ recovery step from the BS-UHF reference, for
raw-pool recovery (no seed-CISD injection) at $d=2\times10^{8}$, $5\times10^{8}$, $10^{9}$,
$4\times10^{9}$ (unboosted, a purely cosmetic ablation trajectory for this dimension, distinct from
the boosted $d=4\times10^{9}$ point used in the extrapolation fit of Fig.~\ref{fig:fe4s4-extrap}),
$8\times10^{9}$, $2\times10^{10}$, and $3\times10^{10}$. Horizontal lines mark BS-UHF
($-327.081$~Ha), BS-UCCSD ($-327.209$~Ha), BS-UCCSD(T) ($-327.222$~Ha), two classical HCI references
($-327.2207693$~Ha at $\varepsilon=2\times10^{-5}$~Ha and $-327.2275123632984$~Ha at $\varepsilon=7\times10^{-6}$~Ha), and the
finite-$D=4000$ DMRG value ($-327.239$~Ha, rounded). $d=8\times10^{9}$ (cyan) reaches
$-327.231551$~Ha at step 98; from step 78
onward, a same-spin recombination candidate booster (Appendix~\ref{app:recombboost}) is applied on
top of this trajectory (magenta, inset), improving it by a further $\sim$0.1--0.8~mHa to
$-327.232351$~Ha at step 134.
$d=2\times10^{10}$ (dark green) reaches $-327.234400$~Ha at step
85, an operationally converged value under our stopping criterion (per-step gains had fallen to
$\sim$0.008~mHa/step, and the
trajectory was stopped there), and $d=3\times10^{10}$ (orange) reaches
$-327.235076$~Ha at step 73, at which point the trajectory was judged saturated (per-step gains below $0.015$~mHa/step) and stopped.}
\label{fig:master}
\end{figure*}

\begin{table}[t]
\centering
\caption{$[4\mathrm{Fe\text{-}4S}]$ reference energies (Ha) and mean-field $\langle\hat
S^2\rangle$ for the CAS(54e,36o) active space discussed in this section, on the updated, localized
\texttt{FCIDUMP} used throughout \S\ref{sec:results} onward. RHF is the closed-shell restricted
Hartree--Fock energy on this same \texttt{FCIDUMP} (directly computed, not carried over from the
earlier, non-localized \texttt{FCIDUMP}'s value). HCI is variational heat-bath CI from
the BS-UHF determinant, no perturbative correction; the looser (smaller-$\varepsilon$) cutoff is
converged at its own threshold but is not the FCI limit. BS-SQD (deepest point)
is this work's deepest directly computed recovery point; BS-SQD ($d\to\infty$ extrapolated) is the
Gaussian-in-$\log d$ fit of \S\ref{sec:composition}.}
\label{tab:fe4s4-refs}
\begin{tabular}{lcc}
\hline\hline
Method & $E$ (Ha) & $\langle\hat S^2\rangle$\\
\hline
RHF & $-326.5474$ & $0.000$\\
UHF, BS & $-327.0809$ & $8.877$\\
UCCSD, BS & $-327.209$ & ---\\
UCCSD(T), BS & $-327.222$ & ---\\
HCI, $\varepsilon=2\times10^{-5}$ & $-327.2207693$ & ---\\
HCI, $\varepsilon=7\times10^{-6}$ & $-327.2275124$ & ---\\
DMRG, finite $D=4000$ & $-327.2393952$ & $0.000$\\
DMRG, $D\to\infty$ extrapolated & $-327.244797$ & $0.000$\\
BS-SQD (this work, deepest point) & $-327.235076$ & ---\\
BS-SQD ($d\to\infty$ extrapolated) & $-327.244468$ & ---\\
\hline\hline
\end{tabular}
\end{table}

The 72 logical qubits (36 $\alpha$- plus 36 $\beta$-spin orbitals, Jordan--Wigner encoded) were
mapped onto \texttt{ibm\_kingston}'s heavy-hex coupling graph, i.e., assigned to specific physical
qubits and connected by specific two-qubit gates on the real device. This assignment was done
using \texttt{ffsim}'s error-aware
LUCJ layout routine~\cite{ffsim2026,ffsim_software}: rather than starting from the device's
connectivity graph and fitting the circuit to it, the routine starts from which qubit pairs the
LUCJ ansatz itself needs to couple and searches for a placement of those pairs onto the device that
avoids its worst-calibrated qubits and gates. Because
heavy-hex hardware has no direct edge between the $\alpha$- and $\beta$-spin qubit chains, the
routine represents each requested inter-spin coupling with an explicit ancilla bridge
qubit and searches for a subgraph isomorphism~\cite{cordella2004vf2} between this ansatz graph and
a version of the real coupling graph pruned of high-readout-error qubits and high-two-qubit-error
edges using the calibration snapshot in effect at submission time; if the full set of requested
couplings cannot be embedded, the lowest-priority couplings are dropped until an embedding is
found. For this circuit, 5 of 9 requested $\alpha$-$\beta$ couplings were realized this way. The
resulting submitted circuit is active on 88 of the device's 156 qubits: the 72 data qubits above,
5 ancilla bridge qubits realizing the retained inter-spin couplings, and 11 further qubits
introduced by ordinary SWAP-based routing to reconcile the chosen embedding with the physical
lattice's connectivity -- qubits that participate in the circuit's execution but never carry a
spin-orbital's own occupation. The full 88-qubit circuit comprises $32{,}835$ gate instructions
($28{,}747$ single-qubit -- $14{,}674$ $\sqrt{X}$, $13{,}983$ $R_z$, $90$ $X$ -- and $3{,}944$
two-qubit \texttt{cz} gates, the device's native entangling gate) at a two-qubit-gate depth of
$276$ and a total depth of $1{,}155$ ($734$ once the effectively instantaneous, virtual-$Z$
$R_z$ gates, along with barriers and measurements, are excluded). Restricting this same accounting
to instructions acting purely within the 72-qubit chemistry-relevant data register -- i.e.\
excluding any instruction touching a bridge or routing qubit -- gives $28{,}400$ single-qubit and
$3{,}708$ two-qubit gates at a depth of $1{,}095$ ($677$ excluding $R_z$/barriers/measurements): the
16 non-data qubits therefore add comparatively little additional circuit depth ($734-677=57$
layers, $\approx8\%$ relative to the data-register-only depth of $677$) relative to their role in
keeping the ansatz's inter-spin coupling
structure intact on hardware with no native $\alpha$-$\beta$ connectivity. The remaining $144$
instructions on the full 88-qubit circuit not accounted for above are measurement, reset, and other
non-gate instructions (i.e.\ $28{,}747$ single-qubit gates plus $3{,}944$ two-qubit gates plus
$144$ such instructions gives the $32{,}835$ total quoted above).

The deep-recovery campaigns reported below ran on two distinct GPU clusters, Miyabi-G and ROQUO
(specifications and further discussion in \S\ref{sec:hpc}). Both systems run the
\texttt{qcsc-prefect} workflow's classical recovery/diagonalization stage identically; the distinguishing
factor for the results below is simply which cluster a given $d$ was run on and when, not any
difference in the SQD algorithm itself.

On the full $[4\mathrm{Fe\text{-}4S}]$ cluster (54 electrons, 36 orbitals), we sampled
5 million shots on \texttt{ibm\_kingston} of this one-layer LUCJ circuit (DD-XY4 dynamical
decoupling plus measurement-twirling error mitigation) defined by the updated active-space
\texttt{FCIDUMP} and the pairing-$(1,2)$ BS-UHF reference of \S\ref{sec:sweep}, of which reset
mitigation retained $1{,}996{,}334$ shots ($\sim$40\%), for a total quantum processing unit (QPU)
usage time of $\sim$25~min. As with $[2\mathrm{Fe\text{-}2S}]$ (\S\ref{sec:fe2s2}), this
reset-mitigated count is prior to configuration recovery, which is what bridges the gap to the
target particle number. As a pipeline sanity check before any
recovery, we identified the single bitstring measured most frequently across the sampled shots --
a specific occupation pattern, not necessarily identical to the BS-UHF occupation pattern the
circuit was initialized from, since sampling noise can shift which single configuration is most
probable -- and evaluated $\langle x|\hat H|x\rangle$ for that configuration alone. This reproduces
the classical BS-UHF reference to within approximately
$1.4$~mHa
($E=-327.0823$~Ha for the most-probable sampled configuration vs.\ $-327.0809$~Ha for the BS-UHF
determinant), confirming that the sampled distribution is concentrated near the intended reference
without requiring the two bitstrings to coincide exactly. We reused this pool (zero additional
hardware calls) for a $d=10^{8}$ ablation on ROQUO GPUs comparing raw-pool recovery against seeding
the recovery prior with classical singles or singles+doubles configuration-interaction strings
(Table~\ref{tab:seed-cisd-ablation}): seeding with the full S+D manifold at mixing fraction
$f=0.5$ already closes $61\%$ of the remaining gap to BS-UCCSD(T) in a single pass, while
seeding with singles alone gives no gain over raw-pool recovery -- singles alone are not
informative enough to redirect the sampler, while S+D already captures most of the recoverable
correlation this deep-recovery target has not yet reached. This seeding procedure is distinct
from extended SQD~\cite{barison2025extsqd}, which explicitly applies excitation operators to
configurations already present in the SQD subspace; here we instead mix externally supplied
classical CI strings into the recovery prior, without generating new configurations by acting on
existing ones.

\begin{table}[t]
\caption{$[4\mathrm{Fe\text{-}4S}]$, $d=10^{8}$ on ROQUO GPUs: raw-pool recovery vs.\ seeding the
recovery prior with classical CI strings, one recovery pass, mixing fraction $f=0.5$ where
applicable. Gap is measured to BS-UCCSD(T)$=-327.222$~Ha. The raw-pool row is a single-pass
snapshot, not a converged floor: continued raw-pool recovery with no CI seeding at all, carried to
much larger $d$ (Fig.~\ref{fig:master}), subsequently reaches $-327.234400$~Ha ($d=2\times10^{10}$)
and $-327.235076$~Ha ($d=3\times10^{10}$) -- both already deeper than the one-pass
seed-CISD shortcut below, and within approximately $5.0$ and $4.3$~mHa, respectively, of the
finite-$D=4000$ DMRG value. Seeding with classical correlation shortens the number of
raw-pool recovery steps needed to reach a given energy, but is not required for raw-pool recovery
to eventually surpass it.}
\label{tab:seed-cisd-ablation}
\footnotesize
\setlength{\tabcolsep}{0pt}
\renewcommand{\arraystretch}{1.0}
\makebox[\linewidth][l]{%
\begin{tabular}{lccc}
\hline\hline
Run & Seed & $E$ (Ha) & $E-E_{\mathrm{UCCSD(T)}}$\\
\hline
raw pool (1 pass, $d=10^{8}$) & --- & $-327.124$ & $+98$~mHa\\
seed singles ($f{=}0.5$) & S & $-327.117$ & $+105$~mHa\\
seed S+D ($f{=}0.5$) & S+D & $-327.1835$ & $+38$~mHa\\
\hline
raw pool, saturated ($d=2{\times}10^{10}$) & --- & $-327.2344$ & $-12$~mHa\\
\hline\hline
\end{tabular}%
}
\end{table}

At the larger $d=8\times10^{9}$ scale used for the main result, these same reused sample pools
($8\times8$ Davidson communicator grid, 64 nodes) were recovered for 98 recovery steps on
Miyabi-G, with per-step checkpointing. Figure~\ref{fig:master} shows the resulting trajectory:
monotonic convergence crossing BS-UCCSD ($-327.209$~Ha) at step 13 and BS-UCCSD(T)
($-327.222$~Ha) at step 24, reaching $E=-327.231551$~Ha at step 98 ($7.8$~mHa above the
finite-$D=4000$ DMRG value), against a reference-Hamiltonian span of $158$~mHa between BS-UHF and
DMRG. From step 78 onward, this trajectory is continued with a same-spin recombination candidate
booster (Appendix~\ref{app:recombboost}) that synthesizes new, higher-rank candidate determinants
by recombining the excitation patterns of already-sampled, orbital-disjoint low-rank determinants,
reusing information already in the sampled pool rather than drawing additional quantum samples --
conceptually related to, but mechanistically distinct from, extended
SQD~\cite{barison2025extsqd}'s explicit application of excitation operators to configurations
already in the subspace: the booster recombines pairs of already-sampled low-rank determinants
rather than exciting individual ones.
Applied this late in an already-mature trajectory, it gives only a small refinement here
(Fig.~\ref{fig:master}), though it is more effective when enabled from the first recovery step
(Appendix~\ref{app:recombboost}). Continuing recovery to substantially larger $d$ on ROQUO and Miyabi-G goes deeper still: at
$d=2\times10^{10}$, recovery crosses BS-UCCSD(T) by step 14 and reaches
$E=-327.234400$~Ha by step 85 -- $5.0$~mHa above the finite-$D=4000$ DMRG value, with per-step gains of
$\sim$0.008~mHa/step at that point; at $d=3\times10^{10}$, recovery reaches
$E=-327.235076$~Ha by step 73, at which point the trajectory was judged saturated (per-step gains below $0.015$~mHa/step) and stopped.

\begin{figure*}[t]
\centering
\makebox[\textwidth][c]{\includegraphics[width=\textwidth]{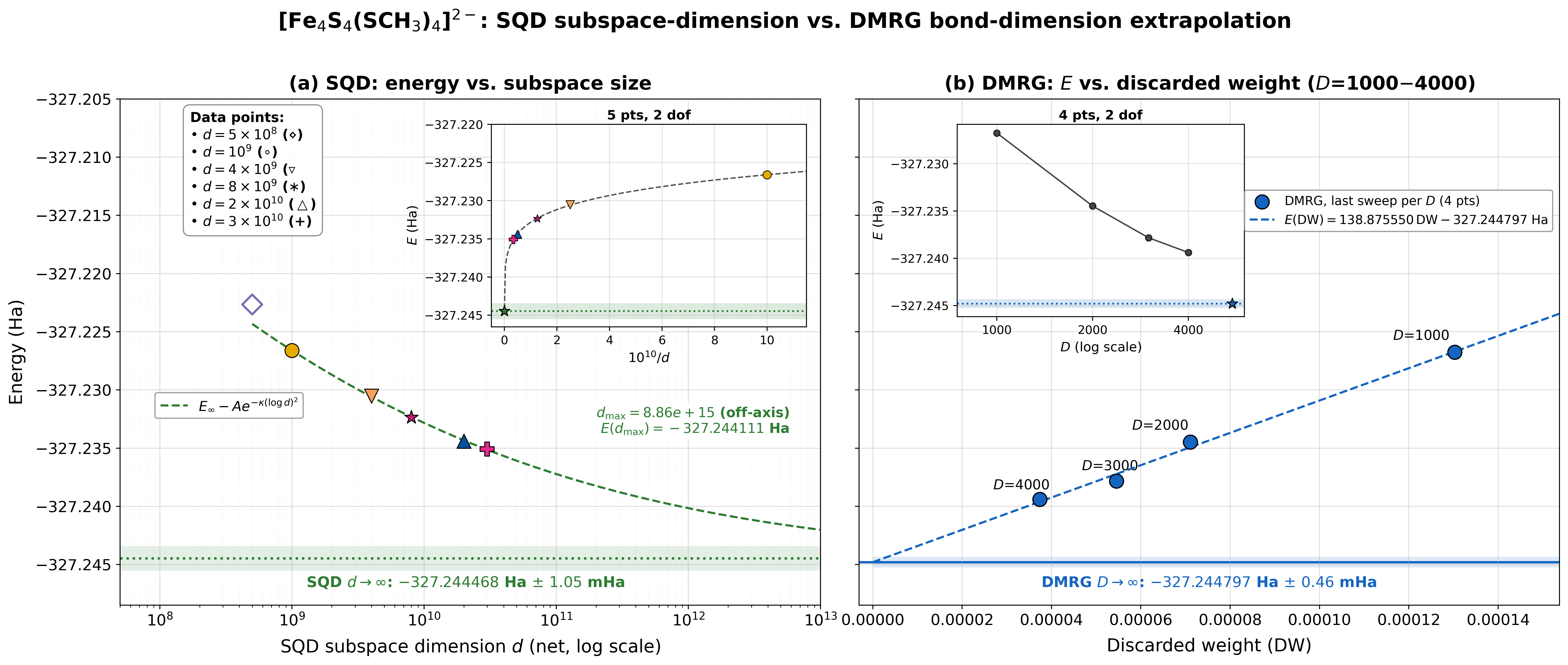}}
\caption{$[4\mathrm{Fe\text{-}4S}]$: two independent extrapolations to the exact-diagonalization
limit for the same active space, analogous to the SQD extrapolation of Fig.~\ref{fig:fe2s2} and
the DMRG cross-check of Fig.~\ref{fig:fe2s2-dmrg} for $[2\mathrm{Fe\text{-}2S}]$, shown here side
by side rather than as separate figures.
\textbf{(a)} SQD energy vs.\ net subspace dimension $d$ (log scale). Fit uses
$d=10^9$, $4\times10^9$, $8\times10^9$, $2\times10^{10}$, and $3\times10^{10}$
(filled markers); $d=2\times10^8$ and $5\times10^8$ (too shallow) are excluded from the fit and
shown as hollow markers. Unweighted least squares. Fit:
$E(d)=E_\infty-A\,e^{-\kappa(\log d)^2}$~\cite{zhai2026femoco} (dashed curve). Shaded band: the
bootstrap $\pm1\sigma$ interval on $E_\infty$ ($\pm1.05$~mHa, 2 degrees of
freedom -- five points, three fit parameters). Inset: the same fit in $10^{10}/d$, where the
$d\to\infty$ intercept (star) is read off. Horizontal lines (main panel only, not shown to keep
the DMRG panel uncluttered): BS-UHF ($-327.081$~Ha) lies above the plotted window; BS-UCCSD
($-327.209$~Ha) and BS-UCCSD(T) ($-327.222$~Ha) fall inside it, near its lower edge.
\textbf{(b)} Independent \texttt{block2}~\cite{block2_2023} DMRG discarded-weight extrapolation
for the same active space: energy vs.\ discarded weight for bond dimensions $D=1000$--$4000$
(circles), linear fit to $D\to\infty$ (dashed line), and the resulting $\pm1\sigma$ band
($\pm0.46$~mHa). The two panels' extrapolated limits (horizontal lines in both panels) differ by
$0.33$~mHa (see text), smaller than their combined $1\sigma$ uncertainty.}
\label{fig:fe4s4-extrap}
\end{figure*}

For this active space, we define the complete-space dimension as
\[
  d_{\max} = \binom{36}{27}^{2} = 8{,}862{,}957{,}169{,}158{,}400 \approx 8.86\times10^{15}.
\]
Thus, even the largest explicitly computed SQD subspace, $d=3\times10^{10}$, spans only
approximately $3.4\times10^{-4}\%$ of the complete space. This fraction is substantially smaller
than the $87.4\%$ reached for $[2\mathrm{Fe\text{-}2S}]$, and consequently the extrapolation to the
complete-space limit is considerably longer range. We extrapolate the five largest subspace
calculations, $d=10^{9}$, $4\times10^{9}$, $8\times10^{9}$, $2\times10^{10}$, and
$3\times10^{10}$, while excluding $d=2\times10^{8}$ and $5\times10^{8}$ as too shallow for the
asymptotic regime. Using the same functional form employed for $[2\mathrm{Fe\text{-}2S}]$ and an unweighted least-squares fit, we obtain $E_{\infty}^{\mathrm{SQD}} = -327.244468 \pm 1.05~\mathrm{mHa}$, as shown in Fig.~\ref{fig:fe4s4-extrap}(a). For clarity, we distinguish three DMRG quantities throughout the following discussion: (i) the finite-bond-dimension result, $E_{D=4000}=-327.2393952$~Ha; (ii) the rounded value $-327.239$~Ha, used only as the $D=4000$ horizontal reference line in Fig.~\ref{fig:master}; and (iii) the discarded-weight-extrapolated DMRG estimate, $E_{D\rightarrow\infty}^{\mathrm{DMRG}} = -327.244797 \pm 0.46~\mathrm{mHa}$, obtained using the spin-adapted \texttt{block2} implementation with an SU(2) bond dimension and the same discarded-weight extrapolation procedure used for $[2\mathrm{Fe\text{-}2S}]$ [Fig.~\ref{fig:fe4s4-extrap}(b)]. The extrapolated SQD and DMRG limits differ by only $0.33$~mHa, which is smaller than their combined $1\sigma$ uncertainty of approximately $1.15$~mHa. Accordingly, the central estimate from each method lies within the $1\sigma$ interval of the other. Finally, using the complete-space dimension $d_{\max}$ defined above, the fitted SQD relation can be evaluated at the complete-space limit, yielding $E(d_{\max})=-327.244111~\mathrm{Ha}$.
This value lies approximately $0.7$~mHa above the discarded-weight-extrapolated DMRG estimate and
is therefore smaller than the conventional $\sim$1.6~mHa (1~kcal/mol) scale, although, as discussed in the Outlook, we do not claim chemical accuracy in the strict sense.
Recent higher-bond-dimension DMRG calculations for closely related $[4\mathrm{Fe\text{-}4S}]$ active-space models predict energies
consistent with this level of accuracy~\cite{arxiv_2603_28648,sharma2014nchem}.

However, DMRG energies for these clusters can be sensitive to how the starting broken-symmetry
reference is prepared because multiple near-degenerate broken-symmetry states are available. Here
we compare against the broken-symmetry manifold used in the earlier literature rather than
re-optimizing a perfectly matched DMRG and SQD reference within a single, unified protocol; our
DMRG calculations were run starting from the RHF FCIDUMP for this active space. At the present scale,
the extrapolated SQD limit already reaches the energy range indicated by recent
higher-bond-dimension DMRG calculations, and the next section shows that increasing $d$ continues
to unlock higher-excitation content beyond what is easily captured at smaller subspaces.

This result is comparable to Kanazawa~et al., who ran the identical closed-loop
architecture on the identical cluster and active-space size (72 spin orbitals) from an
RHF reference at $d=10^{8}$, reporting $E=-326.718$ to $-326.723$~Ha across independent
executions~\cite{kanazawa2025}. Our BS-UHF-referenced result is $512$--$517$~mHa lower --
comparable in magnitude to the entire remaining gap to DMRG in that RHF-based calculation. Becausew
both calculations diagonalize the same Hamiltonian in a subspace of the same recovered-determinant
character, this difference is attributable almost entirely to the mean-field reference the ansatz
and recovery loop were seeded from, not to sampling statistics or subspace size.

\subsection{Subspace composition}
\label{sec:composition}

\begin{figure}[t]
\centering
\makebox[\columnwidth][c]{\hspace*{-0.6em}\includegraphics[width=\columnwidth]{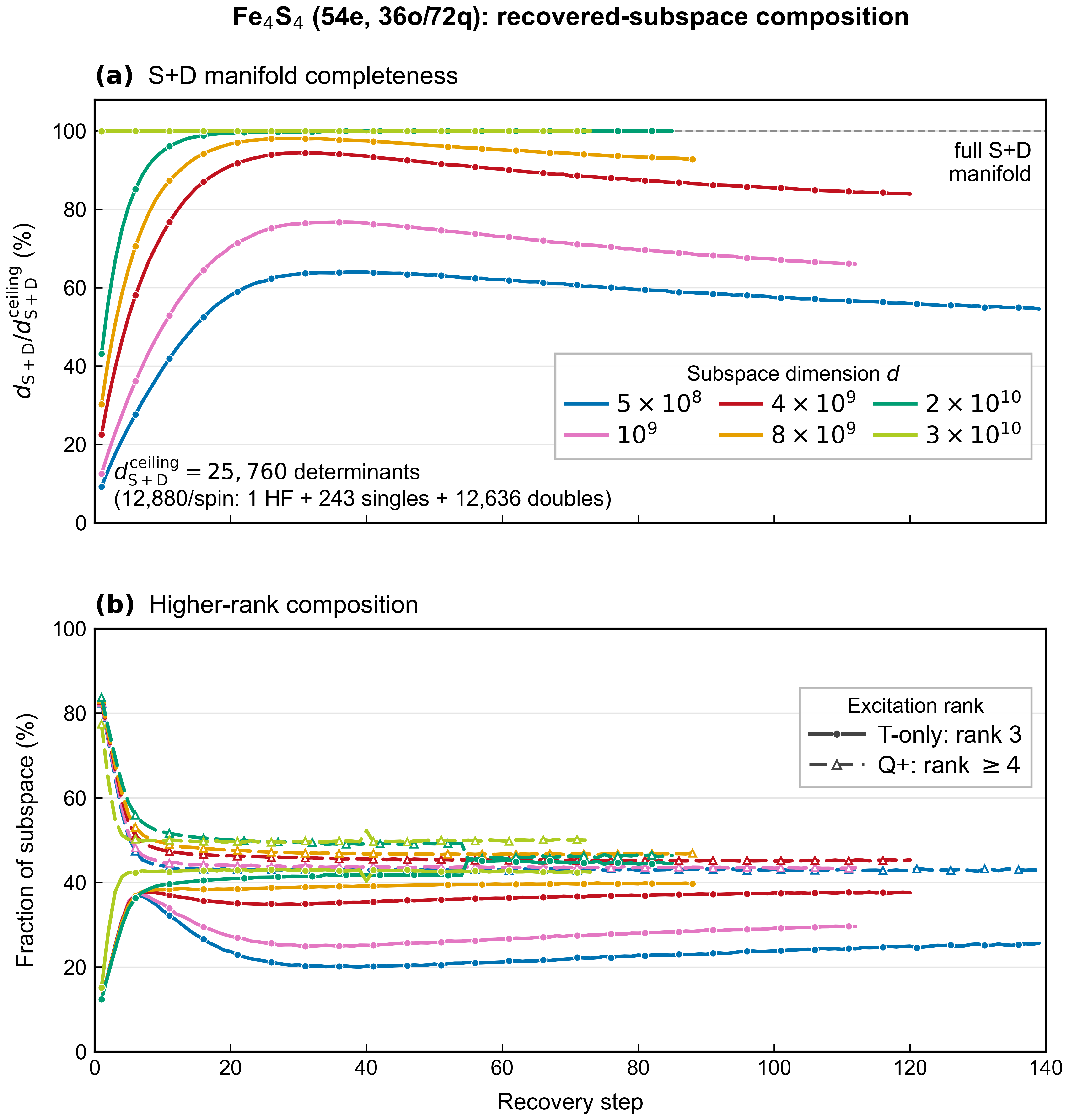}}
\caption{Composition of the recovered determinant subspace by excitation rank, vs.\ recovery
step, across every $d$ we have per-rank data for. Both panels count per-spin (\emph{alpha}-string
or \emph{beta}-string) excitation ranks relative to that spin's own reference string, summed over
both spins -- a different object from the full $(\alpha,\beta)$-determinant-pair total rank used
in Table~\ref{tab:ci-weight}; see \S\ref{sec:composition} for the distinction. \textbf{(a)}
Recovered singles-plus-doubles (S+D) string count as a percentage of the fixed, $d$-independent
S+D ceiling of $25{,}760$ defined in the text, \emph{not} as a fraction of that step's own
(variable) recovered-subspace total -- the latter conflates selection quality with subspace size
and has no fixed reference point across dimensions. The smallest dimension shown reaches a maximum
at little more than half the ceiling and then slowly declines, indicating that some previously
retained S+D strings are replaced during later recovery steps; the manifold is $43\%$ complete at
$d=2\times10^{10}$ and exceeds $99.9\%$ at $d=3\times10^{10}$. \textbf{(b)} The higher-rank
population, as a percentage of that step's own recovered subspace, split into T-only (rank exactly
3, solid) and Q+ (rank $\geq4$, dashed); coarse-format steps with no S/D/T breakdown
($d=2\times10^8$'s steps 1--60) are excluded here. Q+ starts very high ($\sim$80\%, dominated by
sampling noise before recovery has run) and falls as recovery progresses, but a substantial Q+
fraction by count ($\sim$40--50\% by the latest steps across most trajectories) persists rather
than vanishing. This is a statement about per-spin-string counts, not about aggregate wavefunction
weight or energetic importance -- see the weight-resolved discussion and Table~\ref{tab:ci-weight}
below.}
\label{fig:composition}
\end{figure}

Tracking each recovered determinant's excitation rank relative to $|\Phi_0\rangle$
(Fig.~\ref{fig:composition}a) shows why iterative recovery provides information beyond simply
constructing a single initial subspace at fixed $d$: early steps are dominated by replacing
high-excitation-rank noise (rank $\geq5$,
overwhelmingly a sampling artifact) with singles and doubles, consistent with an earlier ablation on
this same cluster in which injecting the full classical singles+doubles manifold
deterministically reached $E=-326.8055$~Ha (against an unpolarized reference) while injecting
thousands of additional sampled higher-rank configurations provided less than $1$~mHa of
additional energy lowering under those settings -- i.e., at that subspace size essentially all
recoverable correlation already lived in singles and doubles. With the BS-UHF reference and a
substantially larger $d$, the recovered subspace approaches this fixed low-rank ceiling, and the
composition of subsequent additions shifts toward triple-and-higher-rank configurations: triples
become the subspace component that continues to grow through step $\sim$24--60 while singles+doubles
saturate, which is consistent
with, but does not by itself establish, an important role for triple-and-higher-rank determinants
in the continued energy improvement beyond the BS-UCCSD(T) benchmark toward DMRG in
Fig.~\ref{fig:master}.

This saturation has a sharp, quantifiable threshold that explains why successive increases in $d$
give very different returns. For this active space, the full singles-plus-doubles manifold has a
fixed size of $12{,}880$ determinants per spin ($1$ Hartree--Fock plus $243$ singles plus $12{,}636$
doubles), i.e.\ $25{,}760$ summed over both spins -- a hard ceiling independent of $d$. Comparing
the very first recovery step across our $d=8\times10^{9}$, $2\times10^{10}$, $3\times10^{10}$, and
$4\times10^{10}$ trajectories against that ceiling: at $d=8\times10^{9}$ the recovered subspace
contains only $7{,}800$ singles-plus-doubles determinants, just $30\%$ of the ceiling; at
$d=2\times10^{10}$ this rises to $11{,}114$ ($43\%$); at $d=3\times10^{10}$ it reaches $25{,}757$,
essentially the full ceiling ($>99.9\%$); and at $d=4\times10^{10}$ it is $25{,}758$ -- already
saturated, with no further room to grow. The first-recovery-step energies themselves are $-327.147313$~Ha ($d=2\times10^{10}$), $-327.219899$~Ha
($d=3\times10^{10}$), and $-327.220081$~Ha ($d=4\times10^{10}$), giving a $72.6$~mHa gain from
$d=2\times10^{10}$ to $d=3\times10^{10}$'s first step and only $0.18$~mHa from $d=3\times10^{10}$ to
$d=4\times10^{10}$'s first step. This behavior suggests that subspace-limited methods may exhibit
non-smooth improvements when an important low-rank string manifold approaches completion -- the
larger jump coincides with the singles-plus-doubles manifold going from $43\%$ to $>99.9\%$ complete,
while the smaller jump adds only
more higher-rank (triples-and-above) determinants on top of an already-complete low-rank manifold --
rather than responding uniformly to increases in nominal subspace size once that
manifold has been completed. The first-step evidence suggests that further dimension increases
beyond $d=4\times10^{10}$ may yield smaller incremental gains than the $d=2\times10^{10}\to
3\times10^{10}$ transition, reflecting a more gradual regime of improvement after low-rank
saturation; a controlled multi-step comparison is required to determine whether this behavior
persists throughout recovery.

Fig.~\ref{fig:composition}b shows that quadruple-and-higher-rank (Q+) determinants remain
numerous in the recovered subspace by count ($\sim$40--50\% of retained determinants, never collapsing toward zero as $d$ grows). Neither DMRG's bond-dimension truncation nor a
quadruples-limited coupled-cluster truncation maps directly onto a maximum determinant excitation
rank -- both methods can represent, or generate, higher-rank determinant content by other means, and
a recent large-scale benchmark of a related iron--sulfur active space~\cite{zhai2026femoco}
confirms that capturing such content explicitly (extrapolating DMRG well past its largest
converged bond dimension, or carrying coupled-cluster theory through quadruples and a partial
accounting of quintuples) is itself expensive by any classical method's own governing resource.
This is a caution against reading rank-cutoff behavior into any single classical method, our own
included.

The count in Fig.~\ref{fig:composition}b establishes rank accessibility, not rank
importance. A determinant appearing in the recovered subspace carries no guarantee about
its wavefunction amplitude: a large Q+ count could coexist with a small aggregate contribution to
the correlation energy, or the specific energetically important Q+ determinants could be
under-represented even while many unimportant ones are retained. Distinguishing these
possibilities requires rank-resolved observables beyond the determinant count itself -- the
aggregate weight by rank, $\sum_{I\in\text{rank}}|c_I|^2$, where $c_I$ is SQD's own Davidson
eigenvector coefficient for determinant $I$ in the diagonalized recovered subspace (not to be
confused with HCI's own, separately diagonalized CI vector, discussed below). We extracted this
directly for all six of our SQD dimensions by saving the converged Davidson eigenvector alongside
the recovered determinant basis, and report it in Table~\ref{tab:ci-weight} alongside the
matching HCI breakdown discussed later in this section.

\begin{table}[t]
\centering
\caption{CI weight ($\sum_{I\in\text{rank}}|c_I|^2$, as a percentage of the total normalized
wavefunction, including rank 0) resolved by excitation rank relative to the reference determinant
$|\Phi_0\rangle$, for SQD (recovered-subspace checkpoints underlying Fig.~\ref{fig:composition};
$d\to\infty$ not applicable since the Davidson eigenvector is per finite subspace) and HCI (both
thresholds of Table~\ref{tab:fe4s4-refs}) side by side. $|c_{\Phi_0}|^2$ is $|\Phi_0\rangle$'s own
weight (rank 0); each row's remaining five entries plus $|c_{\Phi_0}|^2$ sum to $1$ to at least
12 significant figures (a direct check on the extraction, not merely an assumption). SQD's rank
here is the total excitation rank of the full $(\alpha,\beta)$-determinant pair (rank$_\alpha$ +
rank$_\beta$), a different, larger-denominator object than the per-spin-string ranks plotted in
Fig.~\ref{fig:composition}; determinant counts by that per-spin-string definition are not repeated
here. Parenthesized values for HCI only are each rank's determinant count as a percentage of that
threshold's own selected space (HCI has no equivalent figure).}
\label{tab:ci-weight}
{\small
\begin{tabular}{llccccc}
\hline\hline
Method & \shortstack{Dimension\\threshold} & $|c_{\Phi_0}|^2$ & S & D & T & Q+ \\
\hline
SQD & $d=5\times10^{8}$ & $0.488$ & $6.8\%$ & $19.7\%$ & $9.6\%$ & $15.2\%$\\
SQD & $d=1\times10^{9}$ & $0.438$ & $7.8\%$ & $14.5\%$ & $19.2\%$ & $14.8\%$\\
SQD & $d=4\times10^{9}$ & $0.389$ & $8.1\%$ & $16.2\%$ & $18.8\%$ & $18.0\%$\\
SQD & $d=8\times10^{9}$ & $0.358$ & $\phantom{0}9.1\%$ & $15.2\%$ & $15.2\%$ & $24.7\%$\\
SQD & $d=2\times10^{10}$ & $0.341$ & $\phantom{0}9.3\%$ & $14.3\%$ & $15.7\%$ & $26.5\%$\\
SQD & $d=3\times10^{10}$ & $0.333$ & $\phantom{0}6.4\%$ & $16.6\%$ & $15.3\%$ & $28.3\%$\\
\hline
HCI & $\varepsilon=2\times10^{-5}$ & $0.496$ & \shortstack{$15.3\%$\\($0.01\%$)} & \shortstack{$26.8\%$\\($2.13\%$)} & \shortstack{$5.7\%$\\($36.6\%$)} & \shortstack{$2.7\%$\\($61.3\%$)}\\
HCI & $\varepsilon=7\times10^{-6}$ & $0.430$ & \shortstack{$16.9\%$\\($0.002\%$)} & \shortstack{$27.9\%$\\($0.37\%$)} & \shortstack{$8.0\%$\\($16.7\%$)} & \shortstack{$4.1\%$\\($82.9\%$)}\\
\hline\hline
\end{tabular}
}
\end{table}

$|c_{\Phi_0}|^2$ falls monotonically across all six SQD dimensions
($0.488\to0.438\to0.389\to0.358\to0.341\to0.333$):
the reference determinant's own weight in the recovered wavefunction shrinks steadily as $d$
grows, consistent with deeper recovery continuing to redistribute weight onto other
determinants rather than saturating around the seed. Combined T+ (T and Q+) weight rises from
$24.7\%$ to $43.6\%$ over the same range, monotonically, while combined S+D weight
does not move in a single direction ($26.4\%\to22.3\%\to24.3\%\to24.3\%\to23.6\%\to23.1\%$) -- together these show
directly that the persistent high-rank determinant count reported above is accompanied by a
substantial and, for T+, steadily growing aggregate wavefunction weight, not merely numerous
determinants with negligible aggregate weight. What the count alone does establish is that SQD's
recovery procedure imposes no
explicit excitation-rank
cutoff: Q+ determinants can be sampled and retained at every $d$ we have run. This property should
not be interpreted through a direct analogy with DMRG bond dimension, because finite-bond-dimension
MPS wavefunctions can also contain high-excitation-rank determinants relative to a selected
reference.

As an independent classical comparison, we performed variational heat-bath configuration interaction (HCI) calculations for $[\mathrm{Fe}_4\mathrm{S}_4(\mathrm{SCH}_3)_4]^{2-}$ starting from the single BS-UHF determinant~\cite{holmes2016hci}. External determinants $|\Phi_A\rangle$ were selected when $|H_{AI}c_I|>\varepsilon$ for at least one determinant $|\Phi_I\rangle$ in the current wavefunction, with $H_{AI}=\langle\Phi_A|\hat H|\Phi_I\rangle$, followed by rediagonalization (Davidson tolerance $10^{-5}$; no second-order perturbative correction). At $\varepsilon=2\times10^{-5}$~Ha the selected space contained $3{,}946{,}437$ determinants and gave $E=-327.2207693$~Ha; loosening to $\varepsilon=7\times10^{-6}$ yielded $22{,}685{,}530$ determinants and $E=-327.2275123632984$~Ha. The $\varepsilon=2\times10^{-5}$ HCI energy lies $1.2$~mHa above the rounded BS-UCCSD(T) reference of $-327.222$~Ha, whereas the looser-threshold HCI energy lies $5.5$~mHa below it; our deepest SQD point ($d=3\times10^{10}$, step 73, $-327.235076$~Ha) lies $7.6$~mHa below the looser HCI point, while SU(2) DMRG starting from the RHF FCIDUMP for this active space reaches $-327.2393952$~Ha at $D=4000$. HCI's own CI vector -- a distinct object from SQD's Davidson eigenvector above, obtained from HCI's own perturbative-importance-selected subspace rather than SQD's occupancy-recovered one -- supplies the same rank-resolved weight breakdown for HCI, reported for both thresholds in the lower half of Table~\ref{tab:ci-weight} alongside SQD's own values.

Placed side by side in Table~\ref{tab:ci-weight}, SQD's and HCI's weight distributions differ
sharply despite both showing the
same qualitative count-vs-weight pattern (high-rank determinants dominate by count, contribute
less by weight, in both methods). HCI's perturbative threshold suppresses Q+ weight to
$2.7\%$--$4.1\%$ even as Q+ determinants grow from $61.3\%$ to $82.9\%$ of its selected space by
count between the two thresholds -- loosening $\varepsilon$ admits vastly more Q+ determinants
by count while barely moving their aggregate weight, the clearest direct evidence that HCI's
selection is explicitly importance-filtered through $|H_{AI}c_I|$ rather than based primarily on
observed sampling frequency. SQD's Q+ weight, by
contrast, is already comparable to or larger than either HCI value at every dimension tested
($15.2\%$--$28.3\%$), and grows rather than shrinks as $d$
increases. This is consistent with the two methods' selection criteria differing in kind, not
just in degree: HCI admits a determinant according to a perturbative importance measure,
$|H_{AI}c_I|$, exceeding $\varepsilon$, whereas SQD's
occupancy-frequency-based recovery (\S\ref{sec:fe2s2}) selects determinants by how often
compatible configurations are sampled, with no explicit weight or importance criterion at all.
The much larger aggregate Q+ weight in SQD indicates that the two methods produce substantially
different rank-resolved wavefunction distributions. Determining whether SQD assigns weight to
specific determinants excluded by HCI's own selection requires a direct overlap analysis of the
two methods' selected determinant sets; whether any such material is helping SQD's energy
converge, merely inflating the subspace without corresponding energetic benefit, or some mixture
of both is not resolved by the weight data alone and would additionally need the rank-resolved
Hamiltonian-matrix-element or post-convergence ablation evidence described above.

This reframes subspace size $d$ and selection quality as two coupled, not independent, control parameters
for SQD accuracy. At fixed $d$, better selection should raise the fraction of the retained space
occupied by determinants with substantial amplitude weight rather than same-rank sampling noise;
this selection quality is itself set by the circuit and hardware noise, which govern which
determinants are sampled in the first place and are unaffected by $d$. Increasing $d$ instead acts
downstream of sampling, at the occupancy-subsampling step (\S\ref{sec:fe2s2}) that selects a
fixed-size subspace from an already-sampled candidate pool: a larger $d$ lets more of the
already-sampled determinants that matter survive this selection rather than being discarded purely
for lack of room, while also (which our own trajectories keep improving without saturating, per
\S\ref{sec:results}) enlarging the
subspace's raw expressive capacity. Our own data links these two pictures directly: the
first-step completion threshold identified above shows that $d$'s benefit is not
smooth in general
-- it is punctuated by exactly when a given excitation rank's manifold is completed -- while the
persistent, non-collapsing Q+ count of Fig.~\ref{fig:composition}(b) shows that higher ranks remain
accessible throughout, so
larger $d$ keeps offering new higher-rank structure to select from rather than running out of
room. Advancing both control parameters simultaneously (larger $d$ plus improved selection) provides a concrete route
for SQD to approach the full-CI variational limit of the active-space Hamiltonian. This is
conceptually analogous to improving DMRG through larger bond dimensions, appropriate symmetry
adaptation, optimized orbital ordering, and controlled extrapolation, although the governing
truncation criteria of SQD and DMRG are fundamentally
different~\cite{zhai2026femoco}.

\section{Portability across HPC systems: Fugaku, Miyabi-G, and ROQUO}
\label{sec:hpc}

This section's central point is portability: the \texttt{qcsc-prefect} workflow's classical
recovery/diagonalization stage was built and run, at different times rather than concurrently, on
three architecturally distinct HPC systems -- Fugaku, Miyabi-G, and ROQUO -- introduced once here
and referred to by name only, without repeating hardware specifications, for the remainder of the
paper. Fugaku is RIKEN R-CCS's CPU-only system (A64FX
processors). Miyabi-G is
JCAHPC's GH200-based system, with 1{,}120 compute nodes each built around a single NVIDIA GH200
Grace--Hopper superchip~\cite{jcahpc_miyabi}. ROQUO is RIKEN R-CCS's newer,
quantum-integrated GPU supercomputer at Kobe: 135 nodes in an NVIDIA GB200 NVL4 configuration (4
NVIDIA Blackwell GPUs and 2 NVIDIA Grace CPUs per node, 540 GPUs total), purpose-built as the
classical-HPC side of Japan's quantum--HPC hybrid platform, with direct connectivity to
Fugaku and to IBM's quantum systems~\cite{riken_roquo}. All results elsewhere in this work were
produced by a single system at a time, not by any simultaneous or split multi-system calculation;
each reported energy is traceable to one specific run on one specific cluster.

Most of the recovery runs reported in this work were executed on ROQUO. The
additional runs on Miyabi-G and Fugaku were carried out not for any
single headline energy but primarily to validate the \texttt{qcsc-prefect} workflow itself: to
confirm that the same solver source and the same recovery/diagonalization algorithm build and run
correctly across architecturally distinct systems, that a checkpointed run's accuracy is
unaffected by which system produced it, and that a checkpoint (occupancies and carryover set) can
be transferred between systems and resumed without
any change to the underlying algorithm. ROQUO, as the newest and least-tuned of the three
deployments, is also where we look for evidence of how much headroom remains for
site-specific tuning. Table~\ref{tab:sizing} summarizes, for every ROQUO run to date, the
communicator grid sizing (how $d$ maps to determinants/spin and to the MPI rank grid), the node
count, and the resulting per-step Davidson-solve compute time. The $d=8\times10^{9}$ result of
Fig.~\ref{fig:master} ran on 64 Miyabi-G nodes over 98 checkpointed recovery steps.

\begin{table}[t]
\centering
\caption{ROQUO communicator grid sizing and per-step Davidson-solve compute time,
$\mathrm{adet}\times\mathrm{bdet} =$ total MPI ranks, dets/spin $\approx\sqrt{d}$ (rounded).
Because all recovery steps execute within a single allocation (one \texttt{sbatch} submission
reserves nodes for the whole multi-step run), there is no additional scheduler wait between steps;
the initial job's own queue time is excluded from the reported timings. All ROQUO timings in this table and in the text below were measured between 20 July and 21 August 2026 (job-accounting start/end times); ROQUO's software stack and the \texttt{sbd} library have been updated since, so these figures may change with later experiments. Every row uses the
same-spin recombination booster of Appendix~\ref{app:recombboost} enabled throughout, which makes
the recovered determinant set denser at a given step and is therefore somewhat slower per step
than the un-boosted baseline at the same $d$ (Appendix~\ref{app:recombboost} quantifies this
directly, $2.8$--$3.6\times$ slower per step at the two subspace sizes compared there, but far
fewer steps needed to reach a given energy, so faster overall in wall-clock time).}

\label{tab:sizing}
\resizebox{\columnwidth}{!}{%
\begin{tabular}{lccccc}
\hline\hline
$d$ & dets/spin & grid & ranks & nodes & Davidson time/step\\
\hline
$5{\times}10^{8}$ & $2.24{\times}10^{4}$ & $2{\times}2$ & 4 & 1 & 00:11:24--00:11:51\\
$1{\times}10^{9}$ & $3.2{\times}10^{4}$ & $2{\times}2$ & 4 & 1 & 00:13:22--00:13:50\\
$4{\times}10^{9}$ & $6.3{\times}10^{4}$ & $4{\times}4$ & 16 & 4 & 01:16:00--01:17:00\\
$8{\times}10^{9}$ & $8.9{\times}10^{4}$ & $4{\times}8$ & 32 & 8 & 01:54:00--01:59:00\\
$8{\times}10^{9}$ & $8.9{\times}10^{4}$ & $8{\times}8$ & 64 & 16 & 01:54:00--01:58:00\\
$2{\times}10^{10}$ & $1.41{\times}10^{5}$ & $8{\times}16$ & 128 & 32 & 03:44:00\\
$3{\times}10^{10}$ & $1.73{\times}10^{5}$ & $8{\times}18$ & 144 & 36 & 04:57:00--07:20:00\\
\hline\hline
\end{tabular}}
\end{table}

Each entry in the compute column is the wall-clock time to diagonalize one recovery step's
Davidson eigenproblem, not one full recovery cycle: within that time the block-Davidson solver
builds and repeatedly expands a Krylov subspace of block size 20 inside the recovered
determinant subspace, whose size is the product of the two spin-sector string dimensions,
approximately equal to $d$ (e.g.\ $63{,}245^2\approx4\times10^{9}$ for the $d=4\times10^{9}$ row;
the table reports the per-spin dimensions in rounded form), and iterates that block-20 Krylov
expansion until either the residual-vector norm on the lowest Ritz value drops below a fixed
threshold ($10^{-2}$, dimensionless, not an energy tolerance) or a per-step iteration cap is
reached, before the next recovery step's sampling/selection stage begins. At the
largest $d$ tested, the iteration cap is what actually terminates the solve in practice: the
energy still changes by several mHa on the last permitted iteration, with the residual norm still
above threshold. This per-step Krylov-expansion rule is therefore a practical stopping point for
that step's linear algebra, not the source of the sub-mHa per-step energy change used
elsewhere in this work as a saturation criterion (e.g.\ the $<0.017$~mHa/step threshold of
\S\ref{sec:composition} and the explicit per-step gains reported for $d=2\times10^{10}$ and
$d=3\times10^{10}$ in \S\ref{sec:results}) -- that quantity instead comes from the recovery
procedure's own step-to-step energy trend
(\S\ref{sec:results}), i.e., how much the energy still moves from one completed recovery step to
the next. So the same 20-vector Krylov
block is expanded against a subspace that ranges from $5\times10^{8}$ (smallest table row) to
$3\times10^{10}$ determinant-pairs (largest), and the compute-column entry is the resulting
wall-clock cost of driving that expansion at each $d$ and rank/node
configuration, not a fixed number of Krylov iterations -- larger $d$ both enlarges the
matrix-vector product cost per Davidson iteration and, in practice, tends to need more
iterations before hitting the cap, so the two effects compound in the reported times.

Across the runs examined in Table~\ref{tab:sizing}, we generally observe a rising-then-leveling-off
pattern in per-step Davidson time: early recovery
steps are comparatively fast, when the carryover determinant set is still sparse
and dominated by low-rank (singles/doubles) excitations;
as recovery matures the carryover set becomes denser and increasingly populated by higher-rank
excitations, and the per-step time rises before leveling off once the set of
recovered determinants stabilizes. The present measurements establish this empirical trend but do
not isolate excitation rank as its sole cause -- Davidson cost and iteration count can also depend
on matrix connectivity, spectral gap, load balance, and communication volume, none of which we have
separately measured here. This pattern holds independently of $d$ or node count -- a higher-$d$
run is not directly comparable to a lower-$d$ run in absolute per-step time, since it is solving a
larger subspace throughout.

Building and running \texttt{qcsc-prefect}'s classical solver on Fugaku, Miyabi-G, and ROQUO
(introduced above) required a different environment setup on each system, reflecting their different processor
architectures, compilers, and MPI stacks. Fugaku's solver build uses its own
Fujitsu compiler and MPI wrapper and a vendor math library, and is a
CPU/MPI-only binary with no GPU-specific code path at all. Miyabi-G's build uses its own
site-provided, already-tuned NVHPC/MPI toolchain, while
ROQUO's uses a separately built HPC-X OpenMPI
stack, launched via \texttt{srun} with the \texttt{pmix} plugin -- a newer deployment that has not
yet received the same level of site-specific tuning. Runs on all three systems in this work use
the same solver source and the same recovery/diagonalization algorithm; the choice of
compiler/MPI/toolchain module set was site-specific. We did not perform a bitwise or
tolerance-quantified numerical cross-check of energies produced by different systems from the same
checkpoint; what we verified is that a transferred checkpoint resumes successfully and continues
improving under the same recovery procedure (below).

Comparing matched-workload configurations across systems makes the current tuning gap visible for ROQUO gpu-cluster.
At the same rank count (128 ranks) and the same $d=2\times10^{10}$ subspace, transferring a
Miyabi-G checkpoint (occupancies and carryover set, both independent of the communicator grid) to
a ROQUO job configured with the same $8\times16$ rank grid gives a comparison of the two systems'
compute time under a matched algorithmic configuration: the Davidson solver with this
$d=2\times10^{10}$
Krylov subspace and $8\times16$/128-rank grid finishes in 20--21~min on the current Miyabi-G
deployment, versus 3~h~44~min for the identical configuration on the ROQUO deployment as of August 2026 (12--19 August 2026 runs; subject to change as ROQUO and \texttt{sbd} are updated).
Because the software stacks, process placement, GPU binding, compiler settings, and communication
libraries differ between the two deployments, this comparison characterizes the present end-to-end
deployments rather than the underlying hardware in isolation; it is nonetheless consistent with
ROQUO's own deployment needing further site-specific tuning. A qualitative, dimension-matched but
not grid-matched, comparison is also available against Fugaku: on the
CPU-only \texttt{large} resource group, Fugaku solves a $d=4\times10^{9}$ problem using a
$128\times144$/18{,}432-rank
grid across 2{,}304 CPU nodes in 1~h~47~min--1~h~59~min per step; the matching ROQUO
measurement, at the same $d=4\times10^{9}$ subspace on a $4\times4$/16-rank grid across
4 GPU nodes, requires 1~h~16~min--1~h~17~min per step. Because the architecture,
process count, and resource scale all differ, these values demonstrate portability but do not
provide a controlled performance comparison between the two systems. A cleaner, fixed-problem-size
comparison is available on ROQUO alone: at fixed $d=8\times10^{9}$,
doubling ROQUO's own rank count from 32 ranks on 8 nodes to 64 ranks on 16 nodes produces no
measurable reduction in per-step time (1~h~54~min--1~h~59~min vs.\ 1~h~54~min--1~h~58~min),
indicating poor strong-scaling efficiency over this range under the current process placement and
software configuration.

Taken together, these three deployments show that the BS-UHF \texttt{qcsc-prefect} workflow can be
built reproducibly on three architecturally different HPC targets, and that a recovery run's
checkpoints, occupancies, and carryover determinant sets can be transferred from one system to
another and resumed there successfully, under the same recovery algorithm; the timing comparisons
above characterize the current deployments' relative performance and site-specific tuning
headroom, not the underlying hardware capability in isolation.

\section{Conclusion}

Extending \texttt{qcsc-prefect} with a broken-symmetry UHF reference -- an atom-localized
antiferromagnetic initial guess combined with internal-and-external stability-following that
reaches the correct Noodleman sublattice basin -- provides an alternative and substantially
lower-energy starting point
for SQD calculations on $[4\mathrm{Fe\text{-}4S}]$. On the earlier, non-localized FCIDUMP used to
develop the preparation procedure, the broken-symmetry initialization lowers the mean-field energy
by 219~mHa relative to the closed-shell solution reached from the default
energy-ordered guess (the restricted Hartree--Fock molecular-orbital basis of that
\texttt{FCIDUMP}, occupied in canonical energy order rather than by fragment-localized spin
polarization). The production calculations use the deeper pairing-(1,2) BS-UHF solution
obtained on the updated, localized FCIDUMP ($E=-327.0809$~Ha).

Combined with checkpointed, multi-step configuration recovery, the deepest directly computed SQD
trajectory reaches $E=-327.235076$~Ha at $d=3\times10^{10}$ (step 73, where the trajectory was
judged saturated and recovery was stopped; see Fig.~\ref{fig:master}), approximately
$4.3$~mHa above the finite-$D=4000$ DMRG value ($-327.2393952$~Ha). Relative to
the discarded-weight-extrapolated DMRG estimate, the direct SQD point remains
approximately $9.7$~mHa higher, while the fit-conditional SQD $d\to\infty$
extrapolation ($E_\infty=-327.244468\pm1.05$~mHa) is
statistically consistent with the discarded-weight-extrapolated DMRG estimate
($-327.244797\pm0.46$~mHa) under their reported uncertainties. This deepest point is
$512$--$517$~mHa below the $-326.718$ to $-326.723$~Ha range reported for a
same-active-space, same-workflow calculation run from an RHF reference~\cite{kanazawa2025}; because
the two calculations also differ in subspace dimension, recovery depth, sampled pool, and hardware,
this difference should not be attributed exclusively to the mean-field reference without a
controlled like-for-like comparison.

The recovered-subspace composition shows that increasing $d$ and iterating the recovery procedure
do more than simply add arbitrary determinants: the low-rank spin-string population approaches
saturation, while triple-and-higher-rank configurations remain numerous as recovery proceeds. This
behavior is consistent with an increasing role for higher-rank correlation in the continued energy
improvement toward DMRG, although the present count data do not independently establish its
energetic contribution. More broadly, the results demonstrate that reference-state preparation,
determinant selection, subspace size, and recovery depth are coupled factors in quantum-centric SQD
calculations of multi-reference transition-metal clusters.

The unrestriction dilemma raised in the Introduction deserves a direct assessment. The BS-UHF reference used throughout is not a spin eigenstate: its $\langle\hat S^2\rangle=8.877$ reflects substantial spin contamination from the antiferromagnetically coupled, locally high-spin Fe centers~\cite{noodleman1981}, and the spin structure it encodes matches the pairing pattern of the spin-adapted DMRG singlet of Ref.~\cite{sharma2014nchem}. This contaminati
on is therefore the expected consequence of breaking spin symmetry, not a defect of the particular solution. For the ground-state energies targeted in this work it is a tolerable com
promise, because the determinant serves only as a seed---supplying the LUCJ amplitudes and
initializing sampling---and is judged by the correlated energy it enables, such as the $\approx42$~mHa UCCSD(T) lowering obtained on the localized reference, rather than by its own s
pin purity~\cite{schurkus2020}.

This acceptability is scoped, not universal, and it shifts with the application. For observables that depend on the total-spin quantum number itself---spectra, singlet--triplet (ST)
gaps, magnetic exchange couplings, or any other property that distinguishes states of different $S$---an unconverged $\langle\hat S^2\rangle$ is a source of error, and a spin-pure or spin-projected treatment is required~\cite{yamaguchi1988,david2017}. Redox reactions are a potential application that calls for a closer look at spin contamination and correlation, since both the oxidized and reduced clusters admit several competing spin-coupling
patterns whose mean-field ordering can change under correlation~\cite{zhai2026femoco}. Conversely, there are settings where breaking spin symmetry is not merely tolerable but necessary to reach a qualitatively correct description at all, such as frustrated spin systems or
relativistic systems with spin--orbit coupling---in the latter case because spin--orbit coupling mixes states of different $S$, so total spin is no longer an exact quantum number of
the full Hamiltonian in the first place; the broken-symmetry formalism developed here could serve as a starting point for those as well.

We close with a practical caution. Applications such as thermochemistry depend on accurate ground-state energies and, especially, accurate energy \emph{differences} between states -- a substantially more demanding target than a single absolute energy, since systematic errors that are large in absolute terms can still cancel, or can just as easily compound, when two such energies are subtracted. The results above show that a broken-symmetry reference is promising in this regard, closing most of the accessible gap to a near-exact DMRG benchmark on an absolute-energy footing; whether that accuracy carries through to the energy differences a given thermochemical application actually needs is a question we have not directly tested here and flag for future work.

\section*{Data availability}
The FCIDUMP integral files for $[2\mathrm{Fe\text{-}2S}]$ and $[4\mathrm{Fe\text{-}4S}]$ used in
this work -- both the source (RHF/CASSCF-basis) integrals and the broken-symmetry-UHF (BS-UHF)
MO-basis integrals generated from them -- together with the raw quantum-hardware sample pools
underlying the configuration-recovery results ($[2\mathrm{Fe\text{-}2S}]$: \texttt{ibm\_fez},
100{,}000- and 500{,}000-shot pools; $[4\mathrm{Fe\text{-}4S}]$: \texttt{ibm\_kingston}, a merged
$\sim1.8\times10^6$-shot pool) are available on Zenodo at
\href{https://doi.org/10.5281/zenodo.23064186}{10.5281/zenodo.23064186}~\cite{zenodo_dataset}.
The code implementing the BS-UHF reference construction, the SQD configuration-recovery pipeline,
and the classical extrapolations reported here is available at
\url{https://github.com/kunalroy332/qcsc-prefect}.

\begin{acknowledgments}
We acknowledge fruitful discussions with Toshinari
Itoko and Yuto Terashima.
We gratefully acknowledge Akira Naruse for valuable support in the acceleration of SBD.
We thank Fugaku, Miyabi-G, and ROQUO operations teams for compute access. We acknowledge Jae-Ho Cheong at Yonsei University and Ji-Yong Sung at Seoul National University for insightful discussions regarding the practical utility and applications of UHF-SQD. We also extend our sincere thanks to Sangwook Wu and Yeon Gyo Seo at PharmCADD for inspiring insights into the crucial role of broken-symmetry states in iron–sulfur (Fe-S) clusters. We thank Yukio Kawashima, and especially Kevin Sung, for \texttt{ffsim} explanations, functionality, and guidance.
A portion of this work at RIKEN is based on results obtained from Project No.~JPNP20017, supported by the New Energy and Industrial Technology Development Organization (NEDO). 
This work at RIKEN was also supported by JSPS KAKENHI Grants No.~JP21H04446 and No.~JP26K06972; JST COI-NEXT (Grant No.~JPMJPF2221); the Program for Promoting Research on the Supercomputer Fugaku (Grant No.~MXP1020230411) by MEXT, Japan; and the COE research grant in computational science from Hyogo Prefecture and Kobe City through the Foundation for Computational Science. 
T.S., S.Y., and M.S. also acknowledge support from the UTokyo Quantum Initiative and the RIKEN TRIP initiative (RIKEN Quantum and Software for Hybrid System of Quantum and High-performance Computers).   
\end{acknowledgments}

\appendix

\section{Construction of the broken-symmetry reference}
\label{app:bsuhf}

This appendix records the explicit construction summarized in \S\ref{sec:refstate}. The techniques
are standard for producing broken-symmetry UHF solutions; they are collected here so that the
procedure is reproducible, and because the specific parameter choices matter for the pairing sweep
of \S\ref{sec:sweep}.

\subsection{Fragment-localized seed density}

Let the active-space orbitals be partitioned into disjoint fragments $F$ (e.g., one fragment per
metal center, one for a bridging ligand, one for spectator ligands). Each fragment is assigned to
one of three roles: spin-up, spin-down, or closed (doubly occupied); a bridging
fragment that mediates superexchange may instead be marked free (half-filled, unpolarized).
For a fractional polarization amplitude $p\in[0,1]$, the initial spin-density matrix
$D^{(0)}=(D^{\alpha(0)},D^{\beta(0)})$ is built orbital-diagonal within each fragment $F$ containing
orbital $x$:
\begin{equation}
D^{\alpha(0)}_{xx},\,D^{\beta(0)}_{xx} \;=\;
\begin{cases}
\tfrac12(1+p),\;\tfrac12(1-p) & F \in \text{up} \\[2pt]
\tfrac12(1-p),\;\tfrac12(1+p) & F \in \text{down} \\[2pt]
\tfrac12,\;\tfrac12 & F \in \text{free} \\[2pt]
1,\;1 & \text{closed}
\end{cases}
\label{eq:afguess}
\end{equation}
with off-diagonal elements zero. The fragment roles and orbital counts are chosen so that
$\mathrm{Tr}(D^{\alpha(0)})+\mathrm{Tr}(D^{\beta(0)})$ equals the target active-space electron
number and $\mathrm{Tr}(D^{\alpha(0)})-\mathrm{Tr}(D^{\beta(0)})$ gives the desired $M_S$. At $p=1$,
each magnetic orbital on an up fragment is initialized fully $\alpha$-occupied and $\beta$-empty,
while each orbital on a down fragment is initialized fully $\beta$-occupied and $\alpha$-empty
(the maximally localized Noodleman guess); $p<1$ reduces the initial sublattice
polarization and allows stronger coupling between the nominal spin sublattices, producing a softer
initial guess with a smaller $\langle\hat S^2\rangle$ (Table~\ref{tab:pairing-sweep}).
$D^{(0)}$ is used to seed a plain UHF calculation, $|\Phi^{(0)}\rangle = \mathrm{UHF}(D^{(0)})$,
under the molecule's exact one- and two-electron integrals (imported from the target FCIDUMP, not
re-derived from a basis set). Figure~\ref{fig:localization} shows $D^{(0)}$ at $p=1$ for both
clusters studied.

\subsection{Stability-following}

$|\Phi^{(0)}\rangle$ is then refined to a stable local minimum by stability-following. Standard
(internal) following repeatedly diagonalizes the electronic Hessian restricted to
spin-conserving rotations; $\mathrm{mo}^{\downarrow}_{\mathrm{int}}(\Phi^{(k)})$ denotes the
orbitals displaced along the most negative such Hessian eigenmode when one exists (a step whose
scale is set by the Newton solver's own line search below), and if an instability exists, rebuilds the density from the unstable
mode and reconverges (via a second-order/Newton SCF) before repeating:
\begin{equation}
|\Phi^{(k+1)}\rangle = \mathrm{Newton}\!\left[D\!\left(\mathrm{mo}^{\downarrow}_{\mathrm{int}}\big(\Phi^{(k)}\big)\right)\right],
\label{eq:internal}
\end{equation}
terminating when the lowest internal-stability Hessian eigenvalue is nonnegative within the
solver's default tolerance, i.e.\ $\mathrm{mo}^{\downarrow}_{\mathrm{int}}(\Phi^{(k)})=\mathrm{mo}(\Phi^{(k)})$
(no internal instability remains). As noted in \S\ref{sec:refstate}, for
$[4\mathrm{Fe\text{-}4S}]$ this terminates at a first minimum $219$~mHa below RHF
($E=-326.767$~Ha, $\langle\hat S^2\rangle\approx7.6$) that is not the deepest broken-symmetry
minimum located in our search: following the external (spin-symmetry-breaking) instability from
this initialization provides access to a lower-energy basin that is inaccessible to the
internal-only stability procedure from the same starting point. We do not establish that the
distinct Noodleman pairings (Singlet-I/II/III) are separated specifically by that unstable mode, or
that they are exactly symmetry-equivalent; \S\ref{sec:sweep} instead compares them directly via
correlated (BS-UCCSD(T)) energies on the updated FCIDUMP.
We therefore follow both internal and external instabilities, feeding the unstable mode's orbital
coefficients directly to the Newton solver as its next initial guess (not a density rebuilt from
them, which stalls at the same higher basin), and track the lowest-energy iterate seen across up to
ten rounds (fewer if no negative internal or external Hessian eigenvalue remains sooner):
\begin{align}
|\Phi^{(k+1)}\rangle &= \mathrm{Newton}\!\left[\mathrm{mo}^{\downarrow}_{\mathrm{int+ext}}\big(\Phi^{(k)}\big)\right], \label{eq:external}\\
E_{\mathrm{BS\text{-}UHF}} &= \min_{k} \; E\big(\Phi^{(k)}\big).
\end{align}
This reaches $E=-326.801$~Ha on the earlier, non-localized FCIDUMP used to develop the
procedure.\footnote{On this earlier, non-localized FCIDUMP: the specific
Fe1/Fe2-up, Fe3/Fe4-down pairing used throughout this paper was verified against an independent
collaborator reference built from the same FCIDUMP; the mirror-symmetric Fe1/Fe3-up, Fe2/Fe4-down
pairing converges to a shallower basin and requires the fractional-polarization/free-bridging-sulfur
knobs of Eq.~\eqref{eq:afguess} to reach a comparable depth. This is distinct from the pairing sweep
on the updated, localized FCIDUMP reported in Table~\ref{tab:pairing-sweep}, where full-polarization
(1,3) already reaches a comparable depth to (1,2) without those knobs.}

\section{Independent DMRG cross-check for $[2\mathrm{Fe\text{-}2S}]$}
\label{app:fe2s2dmrg}

\begin{figure}[t]
\centering
\includegraphics[width=\columnwidth]{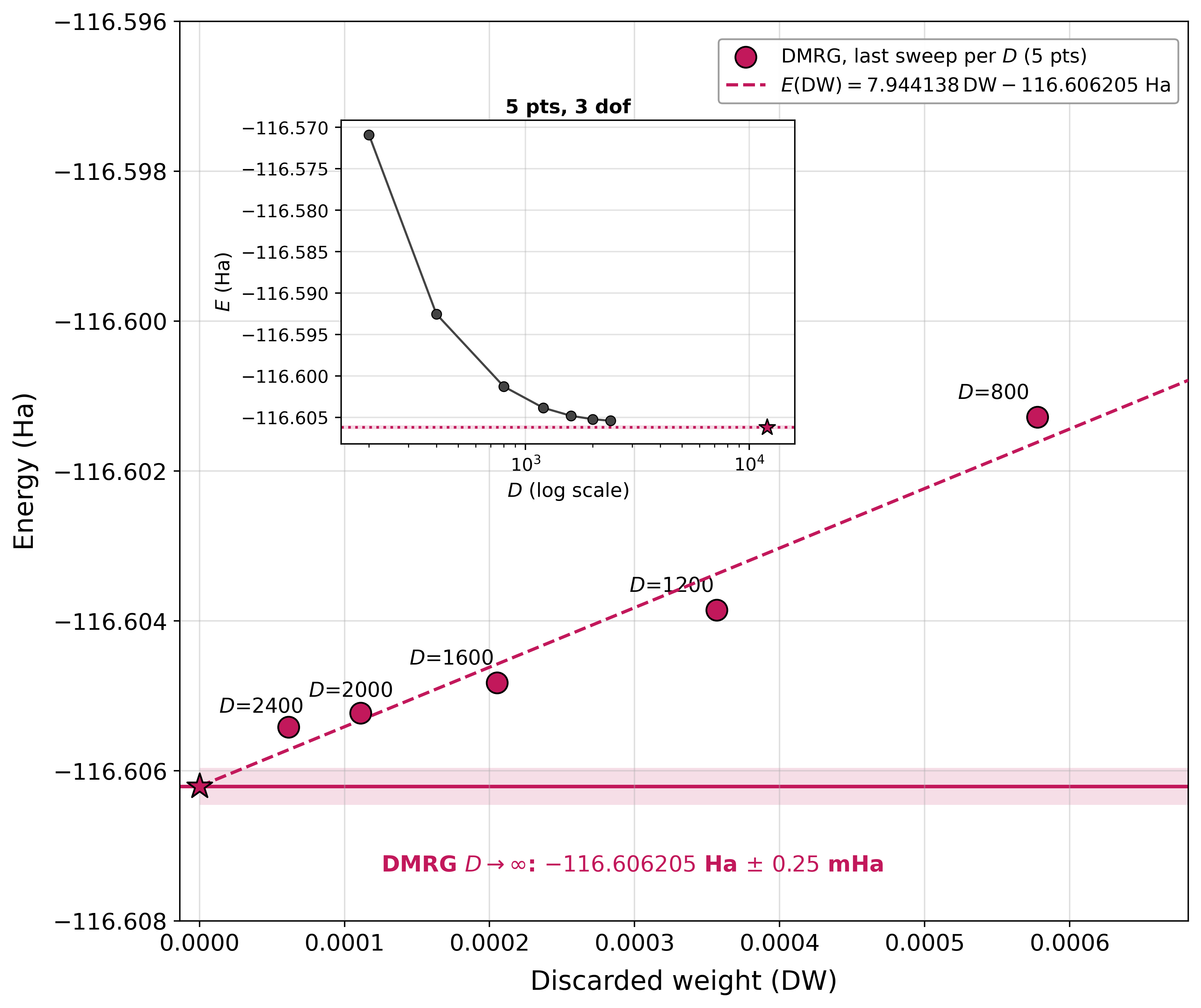}
\caption{$[2\mathrm{Fe\text{-}2S}]$: independent \texttt{block2}~\cite{block2_2023} DMRG
discarded-weight extrapolation for the same active space as Fig.~\ref{fig:fe2s2}: energy vs.\
discarded weight for bond dimensions $D=800$--$2400$ (circles), linear fit to $D\to\infty$
(dashed line), and the resulting $\pm1\sigma$ band ($\pm0.25$~mHa). Inset: the same seven-point
sweep ($D=200$--$2400$) plotted directly against $D$ on a log scale, showing the two points
($D=200,400$) excluded from the linear fit because their discarded weight falls outside the
linear regime of the larger-$D$ points. This DMRG $D\to\infty$ extrapolation differs from the
main text's SQD $d\to\infty$ extrapolation (Fig.~\ref{fig:fe2s2}) by only $0.09$~mHa (computed
from the unrounded fit intercepts; the displayed values round to $0.10$~mHa), each within the
other's $1\sigma$ band -- see \S\ref{sec:fe2s2} for discussion.}
\label{fig:fe2s2-dmrg}
\end{figure}

As discussed in \S\ref{sec:fe2s2}, we cross-checked the SQD subspace-dimension extrapolation of
Fig.~\ref{fig:fe2s2} against an independent DMRG calculation for the same CAS(30e,20o) active
space, using the \texttt{block2} package~\cite{block2_2023} with SU(2)-spin-adapted symmetry.
Fig.~\ref{fig:fe2s2-dmrg} shows the resulting bond-dimension discarded-weight extrapolation.

\section{Recombination-boost timing: early vs.\ mid-trajectory application}
\label{app:recombboost}

\S\ref{sec:results} reports the same-spin recombination candidate booster's effect when switched
on mid-trajectory, on an already-mature $d=8\times10^{9}$ run (branched at step 78; the unboosted trajectory ended at step 98): a small,
$\sim$0.1--0.8~mHa late-stage refinement for negligible additional cost. This appendix reports a
second, separate controlled test of the same booster, switched on from the very first recovery step of a
fresh run, at two smaller subspace dimensions ($d=5\times10^{8}$ and $d=10^{9}$) where an
un-boosted baseline trajectory from an earlier campaign is available for direct comparison.

\begin{figure}[t]
\centering
\includegraphics[width=\columnwidth]{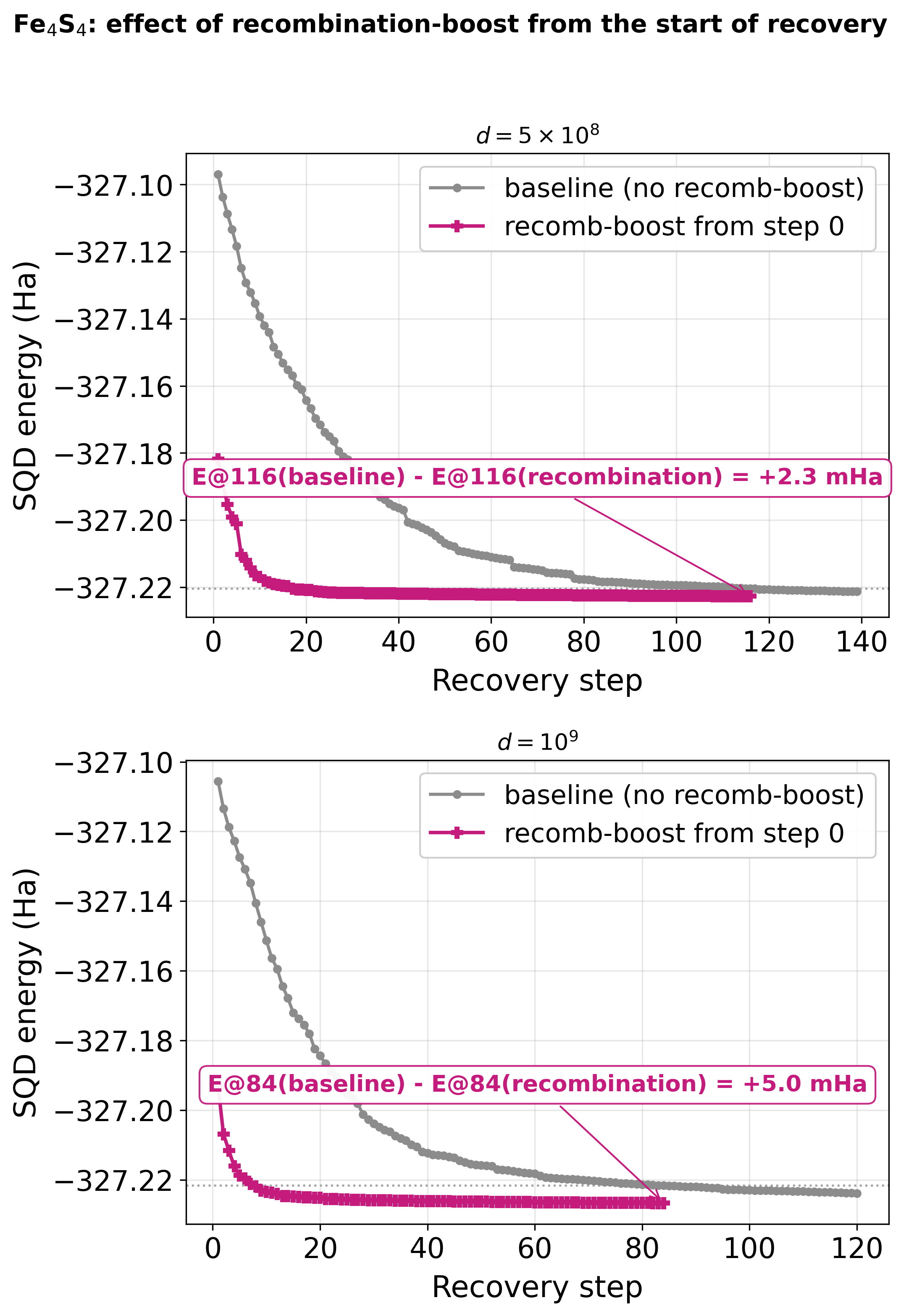}
\caption{Effect of enabling the recombination booster from the first recovery step, for
$d=5\times10^{8}$ (left) and $d=10^{9}$ (right). Gray: baseline raw-pool recovery, no booster.
Magenta: identical run with the booster enabled from step 1. The annotated gap is the energy
difference at the last recovery step common to both trajectories.}
\label{fig:recombboost}
\end{figure}

Both baseline trajectories (no recombination boost) are raw-pool recovery runs from the same
BS-UHF reference and the same sampled pool, each run for 139 ($d=5\times10^{8}$) and
120 ($d=10^{9}$) recovery steps, the point at which each run was stopped; we report the final
energy reached under that stopping point rather than a formally verified convergence criterion for
these two runs specifically. Each baseline was then repeated as a separate controlled test from a fresh
cold start (step 1) with the
booster enabled throughout, on the same hardware and solver build. Figure~\ref{fig:recombboost}
overlays the two pairs of trajectories.

The effect is large at early steps and shrinks steadily as recovery proceeds, consistent with the
booster's own mechanism (\S\ref{sec:results}): it synthesizes new higher-rank candidates by
recombining the excitation patterns of already-sampled low-rank determinants, so its benefit is
largest while the raw sampled pool is still far from converged and most of the useful low-rank
determinants have not yet been found by direct sampling alone. For $d=10^{9}$, the boosted
trajectory is $87.5$~mHa below the baseline at step 1 and $5.0$~mHa below it at step 84, the last
step both trajectories share -- a steady, monotonic narrowing of the gap
rather than a constant offset. The $d=5\times10^{8}$ pair shows the same pattern over a longer
overlap window ($84.9$~mHa at step 1, $12.5$~mHa at step 55, narrowing to $2.3$~mHa at step 116, the last step both trajectories share).

Both boosted runs continued past the point where the two
trajectories still overlap step-for-step, so a sharper comparison is also available: how the
boosted trajectory's own latest energy compares against the baseline's final value
under its own stopping point, and how many boosted steps it took to reach that value. For $d=10^{9}$,
the baseline's final value ($-327.223831$~Ha, its energy after $120$ recovery steps) is
matched by the boosted trajectory at step $12$ -- $10\%$ of the baseline's total step count --
and by its own latest point (step $84$) the boosted trajectory sits $2.8$~mHa below that baseline
final value, not just below the last shared step. The $d=5\times10^{8}$ pair shows the same pattern: the
baseline's final value ($-327.221299$~Ha at step $139$) is matched by the boosted trajectory at
step $23$ ($17\%$ of the baseline's step count), and its latest point (step $116$) sits $1.4$~mHa
below that baseline final value. In both cases the boosted trajectory not only reaches a given energy in fewer steps but,
once run past the baseline's own stopping point, continues descending below the baseline's final
value under its own stopping point.

This comes at a real per-step computational cost: the booster's own candidate-generation and
ranking step adds real wall-clock time to every recovery step, roughly a factor of
$2.8$--$3.6\times$ slower per step than the unboosted baseline at these subspace sizes. The more relevant comparison,
however, is wall-clock time to reach the same energy, not wall-clock time per step: because the
boosted trajectory needs far fewer steps to reach a given energy, it is still substantially faster
overall. At $d=10^{9}$, the boosted run reaches step~1's energy ($-327.193107$~Ha) using $376$~s of
solver wall-clock time; the unboosted baseline does not reach that energy until its own step~24,
after a cumulative $9{,}913$~s -- a $26\times$ reduction in wall-clock time to the same energy. At
$d=5\times10^{8}$, the boosted run's step-1 energy ($-327.181874$~Ha) is not reached by the
baseline until its own step~29 (cumulative $5{,}955$~s), using $372$~s of boosted solver
wall-clock time for step 1, a
$16\times$ reduction. Both comparisons
use only real, measured per-step timings from each trajectory's own job accounting, not an
extrapolation.

Taken together with the mid-trajectory result of \S\ref{sec:results}, this indicates that the
recombination booster is not simply a fixed per-step improvement: its value is concentrated at
early recovery, where the sampled pool is furthest from converged. Its absolute energy advantage
diminishes as recovery proceeds; its late-stage cost-effectiveness is therefore expected to
decrease as well, although a formal marginal-energy-per-wall-clock analysis was not performed.
Enabling it from the start of a run is therefore the more effective use of the same mechanism than
switching it on late, at least for the two subspace sizes tested here ($d=5\times10^8$,
$d=10^9$). The exact protocol differs by dimension: the $d=8\times10^{9}$ trajectory of
\S\ref{sec:results} used mid-trajectory activation at step 78, while the deeper
$d=2\times10^{10}$ and $d=3\times10^{10}$ campaigns used the booster from the first recovery step.
A controlled boosted-versus-unboosted comparison at those larger dimensions, analogous to this
appendix's own $d=5\times10^8$/$d=10^9$ comparison, has not yet been performed.

\section{Outlook: Orbital optimization and parameter tuning}
\label{sec:outlook}

An outer orbital-rotation loop over the recovered subspace was found to give little benefit against
a fixed sample pool: rotating orbitals without re-sampling in the rotated basis actually
raised the recovered energy by $27$~mHa in our tests, from $-326.933$~Ha (before rotation) to
$-326.906$~Ha (after rotation, at matched settings on an
unpolarized reference), because the fixed pool of raw bitstrings is only informative in the basis it
was sampled in. Our results do not show that orbital optimization must occur before an initial
\texttt{FCIDUMP} is generated; rather, they show that post hoc orbital rotation is inconsistent
with reuse of an unchanged sample pool, because the measured bitstrings and recovered coefficients
are basis dependent. A useful orbital-optimization loop therefore needs to transform the
Hamiltonian and regenerate or consistently update the sampled state after each rotation --
consistent
with the differential-evolution outer loop of Ref.~\cite{shirakawa2025} rather than a
post-hoc classical rotation. A second, complementary direction suggested by
\S\ref{sec:composition} is an importance-guided (SHCI/ASCI-style) selection criterion in place of
SQD's purely frequency/occupancy-based recovery, which should accelerate the deep-recovery regime
in which triple-and-higher-rank configurations remain numerous, without simply enlarging $d$. Given
the count-based evidence for higher-rank content in that regime, a further round of sampling guided
by a higher-excitation-aware update -- one that
does not move too far from the previously recovered region of variational-parameter or
wavefunction space -- would
also reduce the recovery effort required to reach a given accuracy. Pursuing this direction is a
step toward tighter agreement with near-exact diagonalization for larger active spaces; the present
deepest direct SQD point
remains several mHa above both the finite-$D$ and extrapolated DMRG references
(\S\ref{sec:results}). We use careful wording here: chemical accuracy properly refers to agreement
between computed and experimental \emph{energy differences} (e.g.\ reaction or excitation
energies), conventionally within about $1$~kcal/mol ($\sim$1.6~mHa), not agreement between a
computed total energy and the exact total energy within a fixed active space. The present
comparison is entirely of the latter kind -- computed vs.\ near-exact total energies within the
same active-space Hamiltonian -- and this work does not examine whether errors in such total
energies cancel in the energy differences that chemical accuracy is actually defined on; we
therefore do not claim chemical accuracy for the results above, only that closing the residual gap
to near-exact diagonalization is a meaningful direction for future work.

\section*{Supporting Information}

\textbf{AF-groups orbital mapping.} For $[4\mathrm{Fe\text{-}4S}]$ (36 orbitals), the fragment map
used throughout is: two spectator-ligand fragments \texttt{l1} (orbitals 0--1) and \texttt{l2}
(34--35); four iron fragments \texttt{fe1} (2--6), \texttt{fe2} (7--11), \texttt{fe3} (24--28),
\texttt{fe4} (29--33); and one bridging-sulfide fragment \texttt{s} (12--23). The pairing used is
$\text{up}=\{\texttt{fe1},\texttt{fe2}\}$, $\text{down}=\{\texttt{fe3},\texttt{fe4}\}$, at full
polarization ($p=1$), which reaches the deepest broken-symmetry basin directly. On the earlier,
non-localized FCIDUMP used to develop the procedure (\S\ref{app:bsuhf}), a mirror-symmetric
$\{\texttt{fe1},\texttt{fe3}\}$/$\{\texttt{fe2},\texttt{fe4}\}$ pairing instead required $p\approx0.6$
together with the bridging-sulfide fragment marked free to reach a comparable depth; on the updated,
localized FCIDUMP used for production (Table~\ref{tab:pairing-sweep}), full polarization alone
already reaches a comparable depth for this pairing.

\textbf{Checkpoint/resume.} Long recovery runs (tens of steps, multi-day wall time on a shared
queue) checkpoint per (trial, walker) -- the outer DE trial and inner walker of
Fig.~\ref{fig:pipeline} -- after every completed recovery step: subspace occupancy,
best energy so far, and the carryover determinant set. Resubmitting the identical launch command
against an existing checkpoint directory auto-skips already-completed steps and continues from the
last saved state; this checkpointing mechanism allowed the 98-step Miyabi-G run in
Fig.~\ref{fig:master} to resume after several
unrelated workflow-orchestrator restarts without loss of completed recovery steps.

\nocite{*}
\bibliographystyle{apsrev4-2}
\bibliography{references}

@article{robledomoreno2025,
  author  = {Robledo-Moreno, J. and Motta, M. and Haas, H.
             and Javadi-Abhari, A. and Jurcevic, P. and Kirby, W.
             and Martiel, S. and Sharma, K. and Sharma, S.
             and Shirakawa, T. and Sitdikov, I. and Sun, R.-Y.
             and Sung, K. J. and Takita, M. and Tran, M. C.
             and Yunoki, S. and Mezzacapo, A.},
  title   = {Chemistry beyond the scale of exact diagonalization
             on a quantum-centric supercomputer},
  journal = {Sci. Adv.},
  year    = {2025},
  volume  = {11},
  number  = {25},
  pages   = {eadu9991},
  eid     = {eadu9991},
  doi     = {10.1126/sciadv.adu9991},
  eprint  = {2405.05068},
  archiveprefix = {arXiv},
  pmcid   = {PMC13109957},
  url     = {https://doi.org/10.1126/sciadv.adu9991}
}

@article{shirakawa2025,
  author        = {Shirakawa, T. and Robledo-Moreno, J. and Itoko, T.
                   and Tripathi, V. and Ueda, K. and Kawashima, Y.
                   and Broers, L. and Kirby, W. and Pathak, H.
                   and Paik, H. and Tsuji, M. and Kodama, Y.
                   and Sato, M. and Evangelinos, C. and Seelam, S.
                   and Walkup, R. and Yunoki, S. and Motta, M.
                   and Jurcevic, P. and Horii, H. and Mezzacapo, A.},
  title         = {Closed-loop calculations of electronic structure on a quantum processor and a classical supercomputer at full scale},
  journal       = {arXiv},
  year          = {2025},
  volume        = {2511.00224},
  eid           = {2511.00224},
  eprint        = {2511.00224},
  archiveprefix = {arXiv},
  primaryclass  = {quant-ph},
  doi           = {10.48550/arXiv.2511.00224},
  url           = {https://arxiv.org/abs/2511.00224}
}

@article{kanazawa2025,
  author        = {Kanazawa, N. and Morohoshi, Y. and Takahashi, H.
                   and Kawashima, Y. and Horii, H. and Nakajima, K.},
  title         = {Observability architecture for quantum-centric supercomputing workflows},
  journal       = {arXiv},
  year          = {2025},
  volume        = {2512.05484},
  eid           = {2512.05484},
  eprint        = {2512.05484},
  archiveprefix = {arXiv},
  primaryclass  = {quant-ph},
  doi           = {10.48550/arXiv.2512.05484},
  url           = {https://arxiv.org/abs/2512.05484}
}

@article{Mende2021,
  author  = {Mende, K. and others},
  title   = {Mitochondrial iron--sulfur clusters: Structure, function, and an overview of their role in human disease},
  journal = {Biochimie},
  volume  = {181},
  pages   = {100--115},
  year    = {2021},
  doi     = {10.1016/j.biochi.2021.11.002},
  url     = {https://www.sciencedirect.com/science/article/pii/S2213231721003244},
  note    = {\url{https://doi.org/10.1016/j.biochi.2021.11.002}}
}

@article{Li2019,
  author  = {Li, Zhendong and Guo, Sheng and Sun, Qiming and Chan, Garnet Kin-Lic},
  title   = {Electronic landscape of the P-cluster of nitrogenase as revealed through many-electron quantum wavefunction simulations},
  journal = {Nature Chemistry},
  volume  = {11},
  number  = {11},
  pages   = {1026--1033},
  year    = {2019},
  doi     = {10.1038/s41557-019-0337-3},
  url     = {https://www.nature.com/articles/s41557-019-0337-3},
  note    = {\url{https://doi.org/10.1038/s41557-019-0337-3}}
}

@article{noodleman1981,
  author  = {Noodleman, Louis},
  title   = {Valence bond description of antiferromagnetic coupling in transition metal dimers},
  journal = {The Journal of Chemical Physics},
  volume  = {74},
  number  = {10},
  pages   = {5737--5743},
  year    = {1981},
  doi     = {10.1063/1.440910},
  url     = {https://pubs.aip.org/aip/jcp/article/74/10/5737/444970/Valence-bond-description-of-antiferromagnetic},
  note    = {\url{https://doi.org/10.1063/1.440910}}
}

@article{sharma2014nchem,
  author  = {Sharma, Sandeep and Sivalingam, Kantharuban and Neese, Frank and Chan, Garnet Kin-Lic},
  title   = {Low-energy spectrum of iron--sulfur clusters directly from many-particle quantum mechanics},
  journal = {Nature Chemistry},
  volume  = {6},
  pages   = {927--933},
  year    = {2014},
  doi     = {10.1038/nchem.2105},
  url     = {https://doi.org/10.1038/nchem.2105},
  note    = {\url{https://doi.org/10.1038/nchem.2105}}
}

@article{yamaguchi1988,
  author  = {Yamaguchi, K. and Jensen, F. and Dorigo, A. and Houk, K. N.},
  title   = {A spin correction procedure for unrestricted {Hartree-Fock} and
             {M{\o}ller-Plesset} wavefunctions for singlet diradicals and polyradicals},
  journal = {Chemical Physics Letters},
  volume  = {149},
  number  = {5-6},
  pages   = {537--542},
  year    = {1988},
  doi     = {10.1016/0009-2614(88)80378-6},
  url     = {https://doi.org/10.1016/0009-2614(88)80378-6},
  note    = {\url{https://doi.org/10.1016/0009-2614(88)80378-6}}
}

@article{neese2009,
  author  = {Neese, Frank},
  title   = {Prediction of molecular properties and molecular spectroscopy with
             density functional theory: From fundamental theory to exchange-coupling},
  journal = {Coordination Chemistry Reviews},
  volume  = {253},
  number  = {5-6},
  pages   = {526--563},
  year    = {2009},
  doi     = {10.1016/j.ccr.2008.05.014},
  url     = {https://doi.org/10.1016/j.ccr.2008.05.014},
  note    = {\url{https://doi.org/10.1016/j.ccr.2008.05.014}}
}

@article{migliore2009,
  author  = {Migliore, Agostino and Sit, Patrick H.-L. and Klein, Michael L.},
  title   = {Evaluation of Electronic Coupling in Transition-Metal Systems Using
             {DFT}: Application to the Hexa-Aquo Ferric--Ferrous Redox Couple},
  journal = {Journal of Chemical Theory and Computation},
  volume  = {5},
  number  = {2},
  pages   = {307--323},
  year    = {2009},
  doi     = {10.1021/ct800340v},
  url     = {https://doi.org/10.1021/ct800340v},
  note    = {\url{https://doi.org/10.1021/ct800340v}}
}

@article{blumberger2015,
  author  = {Blumberger, Jochen},
  title   = {Recent Advances in the Theory and Molecular Simulation of Biological
             Electron Transfer Reactions},
  journal = {Chemical Reviews},
  volume  = {115},
  number  = {20},
  pages   = {11191--11238},
  year    = {2015},
  doi     = {10.1021/acs.chemrev.5b00298},
  url     = {https://doi.org/10.1021/acs.chemrev.5b00298},
  note    = {\url{https://doi.org/10.1021/acs.chemrev.5b00298}}
}

@article{chu2017,
  author  = {Chu, Shibing and Bovi, Daniele and Cappelluti, Francesco and
             Orellana, Alberto Giacomo and Martin, Henry and Guidoni, Leonardo},
  title   = {Effects of Static Correlation between Spin Centers in Multicenter
             Transition Metal Complexes},
  journal = {Journal of Chemical Theory and Computation},
  volume  = {13},
  number  = {10},
  pages   = {4675--4683},
  year    = {2017},
  doi     = {10.1021/acs.jctc.7b00316},
  url     = {https://doi.org/10.1021/acs.jctc.7b00316},
  note    = {\url{https://doi.org/10.1021/acs.jctc.7b00316}}
}

@article{david2017,
  author  = {David, Gr{\'e}goire and Guih{\'e}ry, Nathalie and Ferr{\'e}, Nicolas},
  title   = {What Are the Physical Contents of {Hubbard} and {Heisenberg}
             {Hamiltonian} Interactions Extracted from Broken Symmetry {DFT}
             Calculations in Magnetic Compounds?},
  journal = {Journal of Chemical Theory and Computation},
  volume  = {13},
  number  = {12},
  pages   = {6253--6265},
  year    = {2017},
  doi     = {10.1021/acs.jctc.7b00976},
  url     = {https://doi.org/10.1021/acs.jctc.7b00976},
  note    = {\url{https://doi.org/10.1021/acs.jctc.7b00976}}
}

@article{schurkus2020,
  author  = {Schurkus, Henry F. and Chen, Dian-Teng and Cheng, Hai-Ping and
             Chan, Garnet and Stanton, John F.},
  title   = {Theoretical prediction of magnetic exchange coupling constants from
             broken-symmetry coupled cluster calculations},
  journal = {The Journal of Chemical Physics},
  volume  = {152},
  number  = {23},
  pages   = {234115},
  year    = {2020},
  doi     = {10.1063/1.5144696},
  url     = {https://doi.org/10.1063/1.5144696},
  note    = {\url{https://doi.org/10.1063/1.5144696}}
}

@article{benediktsson2022,
  author  = {Benediktsson, Bardi and Bjornsson, Ragnar},
  title   = {Analysis of the Geometric and Electronic Structure of Spin-Coupled
             Iron--Sulfur Dimers with Broken-Symmetry {DFT}: Implications for
             {FeMoco}},
  journal = {Journal of Chemical Theory and Computation},
  volume  = {18},
  number  = {3},
  pages   = {1437--1457},
  year    = {2022},
  doi     = {10.1021/acs.jctc.1c00753},
  url     = {https://doi.org/10.1021/acs.jctc.1c00753},
  note    = {\url{https://doi.org/10.1021/acs.jctc.1c00753}}
}

@article{barison2025extsqd,
  author  = {Barison, Stefano and Robledo Moreno, Javier and Motta, Mario},
  title   = {Quantum-centric computation of molecular excited states with
             extended sample-based quantum diagonalization},
  journal = {Quantum Science and Technology},
  volume  = {10},
  number  = {2},
  pages   = {025034},
  year    = {2025},
  doi     = {10.1088/2058-9565/adb781},
  url     = {https://doi.org/10.1088/2058-9565/adb781},
  note    = {\url{https://doi.org/10.1088/2058-9565/adb781}}
}

@article{block2_2023,
  author  = {Zhai, Huanchen and Larsson, Henrik R. and Lee, Seunghoon and Cui, Zhi-Hao
            and Zhu, Tianyu and Sun, Chong and Peng, Linqing and Peng, Ruojing
            and Liao, Ke and T{\"o}lle, Johannes and Yang, Junjie and Li, Shuoxue
            and Chan, Garnet Kin-Lic},
  title   = {{Block2}: A comprehensive open source framework to develop and apply
             state-of-the-art {DMRG} algorithms in electronic structure and beyond},
  journal = {The Journal of Chemical Physics},
  volume  = {159},
  number  = {23},
  eid     = {234801},
  year    = {2023},
  doi     = {10.1063/5.0180424},
  eprint  = {2310.03920},
  archiveprefix = {arXiv},
  primaryclass  = {physics.chem-ph},
  url     = {https://doi.org/10.1063/5.0180424},
}

@misc{zdl_fcidump_repo,
  author       = {Li, Z.},
  title        = {Active-space model for iron--sulfur clusters},
  year         = {2026},
  howpublished = {GitHub repository},
  url          = {https://github.com/zhendongli2008/Active-space-model-for-Iron-Sulfur-Clusters},
  note         = {Accessed 16 August 2026}
}

@article{qcsc_2506_20658,
  author        = {Lanes, O. and Beji, M. and C{\'o}rcoles, A. D.
                   and Dalyac, C. and Gambetta, J. M. and Henriet, L.
                   and Javadi-Abhari, A. and Kandala, A. and Mezzacapo, A.
                   and Porter, C. and Sheldon, S. and Watrous, J.
                   and Zoufal, C. and Dauphin, A. and Peropadre, B.},
  title         = {A framework for quantum advantage},
  journal       = {arXiv},
  year          = {2025},
  eprint        = {2506.20658},
  archiveprefix = {arXiv},
  primaryclass  = {quant-ph},
  doi           = {10.48550/arXiv.2506.20658},
  url           = {https://arxiv.org/abs/2506.20658}
}

@article{qcsc_methylene_jctc_2025,
  author  = {Liepuoniute, I. and Doney, K. D. and Robledo-Moreno, J.
             and Job, J. A. and Friend, W. S. and Jones, G. O.},
  title   = {Quantum-centric computational study of methylene singlet and triplet states},
  journal = {J. Chem. Theory Comput.},
  year    = {2025},
  volume  = {21},
  number  = {10},
  pages   = {5062--5070},
  doi     = {10.1021/acs.jctc.5c00075},
  url     = {https://doi.org/10.1021/acs.jctc.5c00075},
  note    = {\url{https://doi.org/10.1021/acs.jctc.5c00075}}
}

@article{prresearch_75pv_hbrx,
  author  = {Mikkelsen, M. and Nakagawa, Y. O.},
  title   = {Quantum-selected configuration interaction with time-evolved state},
  journal = {Phys. Rev. Res.},
  year    = {2025},
  volume  = {7},
  number  = {4},
  pages   = {043043},
  eid     = {043043},
  doi     = {10.1103/75pv-hbrx},
  eprint  = {2412.13839},
  archiveprefix = {arXiv},
  url     = {https://doi.org/10.1103/75pv-hbrx}
}

@article{prresearch_dmn4_snfx,
  author  = {Kanno, K. and Kohda, M. and Imai, R. and Koh, S.
             and Mitarai, K. and Mizukami, W. and Nakagawa, Y. O.},
  title   = {Quantum-selected configuration interaction: Classical diagonalization
             of Hamiltonians in subspaces selected by quantum computers},
  journal = {Phys. Rev. Res.},
  year    = {2026},
  volume  = {8},
  number  = {2},
  pages   = {023268},
  eid     = {023268},
  doi     = {10.1103/dmn4-snfx},
  eprint  = {2302.11320},
  archiveprefix = {arXiv},
  url     = {https://doi.org/10.1103/dmn4-snfx}
}

@article{hpc_sched_2506_10052,
  author        = {Bacher, U. and Birmingham, M. and Carothers, C. D.
                   and Damin, A. and Gonzalez Calaza, C. D.
                   and Karnad, A. K. and Mensa, S. and Moreau, M.
                   and Nober, A. and Ohtani, M. and Rossmannek, M.
                   and Rubin, P. and Sahin, M. E. and Wallis, O.
                   and Shehata, A. and Sitdikov, I. and Wennersteen, A.},
  title         = {Quantum resources in resource management systems},
  journal       = {arXiv},
  year          = {2025},
  eprint        = {2506.10052},
  archiveprefix = {arXiv},
  primaryclass  = {quant-ph},
  doi           = {10.48550/arXiv.2506.10052},
  url           = {https://arxiv.org/abs/2506.10052}
}

@article{future_generation_computer_systems_2024_03583,
  author  = {Beck, T. and Baroni, A. and Bennink, R.
             and Buchs, G. and Coello P{\'e}rez, E. A.
             and Eisenbach, M. and Ferreira da Silva, R.
             and Gopalakrishnan Meena, M. and Gottiparthi, K.
             and Groszkowski, P. and Humble, T. S.
             and Landfield, R. and Maheshwari, K. and Oral, S.
             and Sandoval, M. A. and Shehata, A. and Suh, I.-S.
             and Zimmer, C.},
  title   = {Integrating quantum computing resources into scientific
             {HPC} ecosystems},
  journal = {Future Gener. Comput. Syst.},
  year    = {2024},
  volume  = {161},
  pages   = {11--25},
  doi     = {10.1016/j.future.2024.06.058},
  eprint  = {2408.16159},
  archiveprefix = {arXiv},
  url     = {https://doi.org/10.1016/j.future.2024.06.058}
}

@article{hpc_sched_2412_18519,
  author        = {Mantha, Pradeep and Kiwit, Florian J. and Saurabh, Nishant
                   and Jha, Shantenu and Luckow, Andre},
  title         = {{Pilot-Quantum}: {A} {Quantum}-{HPC} Middleware for Resource,
                   Workload and Task Management},
  journal       = {arXiv},
  year          = {2024},
  eprint        = {2412.18519},
  archiveprefix = {arXiv},
  primaryclass  = {quant-ph},
  doi           = {10.48550/arXiv.2412.18519},
  url           = {https://arxiv.org/abs/2412.18519},
}

@inproceedings{hpc_sched_ccgrid_2025_00070,
  author    = {Mantha, Pradeep and Kiwit, Florian J. and Saurabh, Nishant
               and Jha, Shantenu and Luckow, Andre},
  title     = {{Pilot-Quantum}: {A} Middleware for {Quantum}-{HPC} Resource,
               Workload and Task Management},
  booktitle = {2025 IEEE 25th International Symposium on Cluster, Cloud and
               Internet Computing (CCGrid)},
  year      = {2025},
  pages     = {164--173},
  publisher = {IEEE},
  doi       = {10.1109/CCGRID64434.2025.00070},
  url       = {https://doi.org/10.1109/CCGRID64434.2025.00070},
}

@inproceedings{acm_3731599_3767551,
  author    = {Mansfield, E. and Seegerer, S. and Vesanen, P.
               and Echavarria, J. and Farooqi, M. N. and Mete, B.
               and Schulz, L.},
  title     = {First practical experiences integrating quantum computers
               with {HPC} resources: A case study with a 20-qubit
               superconducting quantum computer},
  booktitle = {Proceedings of the Workshops of the International Conference
               for High Performance Computing, Networking, Storage and
               Analysis},
  series    = {{SC Workshops '25}},
  year      = {2025},
  publisher = {Association for Computing Machinery},
  address   = {New York, NY, USA},
  doi       = {10.1145/3731599.3767551},
  eprint    = {2509.12949},
  archiveprefix = {arXiv},
  url       = {https://doi.org/10.1145/3731599.3767551}
}

@inproceedings{ieee_11018264,
  author    = {Ahmed, H. and Mete, B. and Heller, H. and Tovey, M.
               and Deng, X. and Zulfiqar, A. and Farooqi, M. N.
               and Abuzayed, M. and Schulz, M. and Schulz, L. B.},
  title     = {Telemetry for quantum systems in {HPC} centers},
  booktitle = {ISC High Performance 2025},
  year      = {2025},
  pages     = {1--11},
  publisher = {IEEE},
  eid       = {11018264},
  url       = {https://ieeexplore.ieee.org/document/11018264},
  note      = {\url{https://ieeexplore.ieee.org/document/11018264}}
}

@article{lucj_2511_22476,
  author        = {Lin, W.-H. and Liang, F. and Motta, M.
                   and Zhang, H. and Merz, K. M., Jr.
                   and Sung, K. J.},
  title         = {Improved parameter initialization for the (local)
                   unitary cluster {Jastrow} ansatz},
  journal       = {arXiv},
  year          = {2025},
  eprint        = {2511.22476},
  archiveprefix = {arXiv},
  primaryclass  = {quant-ph},
  doi           = {10.48550/arXiv.2511.22476},
  url           = {https://arxiv.org/abs/2511.22476}
}

@article{rsc_51_5_1659,
  author  = {Anand, A. and Schleich, P. and Alperin-Lea, S.
             and Jensen, P. W. K. and Sim, S. and D{\'i}az-Tinoco, M.
             and Kottmann, J. S. and Degroote, M. and Izmaylov, A. F.
             and Aspuru-Guzik, A.},
  title   = {A quantum computing view on unitary coupled cluster theory},
  journal = {Chem. Soc. Rev.},
  year    = {2022},
  volume  = {51},
  number  = {5},
  pages   = {1659--1684},
  doi     = {10.1039/D1CS00932J},
  eprint  = {2109.15176},
  archiveprefix = {arXiv},
  url     = {https://doi.org/10.1039/D1CS00932J}
}

@article{motta_jctc_2019_ucj,
  author        = {Matsuzawa, Y. and Kurashige, Y.},
  title         = {Jastrow-type decomposition in quantum chemistry for
                   low-depth quantum circuits},
  journal       = {J. Chem. Theory Comput.},
  year          = {2020},
  volume        = {16},
  number        = {2},
  pages         = {944--952},
  doi           = {10.1021/acs.jctc.9b00963},
  eprint        = {1909.12410},
  archiveprefix = {arXiv},
  primaryclass  = {physics.chem-ph},
  url           = {https://doi.org/10.1021/acs.jctc.9b00963}
}

@article{rsc_sc_14_40_11213,
  author  = {Motta, M. and Sung, K. J. and Whaley, K. B.
             and Head-Gordon, M. and Shee, J.},
  title   = {Bridging physical intuition and hardware efficiency for
             correlated electronic states: The local unitary cluster
             {Jastrow} ansatz for electronic structure},
  journal = {Chem. Sci.},
  year    = {2023},
  volume  = {14},
  number  = {40},
  pages   = {11213--11227},
  doi     = {10.1039/D3SC02516K},
  url     = {https://doi.org/10.1039/D3SC02516K},
  note    = {\url{https://doi.org/10.1039/D3SC02516K}}
}

@article{pmc11866757,
  author  = {Blunt, N. S. and Caune, L. and Quiroz-Fernandez, J.},
  title   = {Quantum computing approach to fixed-node Monte Carlo
             using classical shadows},
  journal = {J. Chem. Theory Comput.},
  year    = {2025},
  volume  = {21},
  number  = {4},
  pages   = {1652--1666},
  doi     = {10.1021/acs.jctc.4c01468},
  pmcid   = {PMC11866757},
  pmid    = {39909406},
  url     = {https://doi.org/10.1021/acs.jctc.4c01468},
  note    = {\url{https://doi.org/10.1021/acs.jctc.4c01468}}
}

@article{ccsd_rmp_2007,
  author    = {Bartlett, R. J. and Musia{\l}, M.},
  title     = {Coupled-cluster theory in quantum chemistry},
  journal   = {Rev. Mod. Phys.},
  year      = {2007},
  volume    = {79},
  number    = {1},
  pages     = {291--352},
  publisher = {American Physical Society},
  doi       = {10.1103/RevModPhys.79.291},
  url       = {https://doi.org/10.1103/RevModPhys.79.291},
  note      = {\url{https://doi.org/10.1103/RevModPhys.79.291}}
}

@article{arxiv_2602_01165,
  author        = {McFarthing, S. and Pellow-Jarman, A.
                   and Petruccione, F.},
  title         = {Noise-resilient quantum chemistry with half the qubits},
  journal       = {arXiv},
  year          = {2026},
  eprint        = {2602.01165},
  archiveprefix = {arXiv},
  primaryclass  = {quant-ph},
  doi           = {10.48550/arXiv.2602.01165},
  url           = {https://arxiv.org/abs/2602.01165}
}

@article{arxiv_2603_28648,
  author        = {Legeza, {\"O}. and Menczer, A.
                   and Werner, M. A. and Xantheas, S. S.
                   and Neese, F. and Ganahl, M. and Brower, C.
                   and Rodriguez Bernabeu, S. and Hammond, J.
                   and Gunnels, J.},
  title         = {Hunting for quantum advantage in electronic structure
                   calculations is a highly non-trivial task},
  journal       = {arXiv},
  year          = {2026},
  eprint        = {2603.28648},
  archiveprefix = {arXiv},
  primaryclass  = {physics.chem-ph},
  doi           = {10.48550/arXiv.2603.28648},
  url           = {https://arxiv.org/abs/2603.28648}
}

@article{ffsim2026,
  author        = {Sung, K. J. and Choi, I. and Amico, M. and Andrews, B.
                   and Ayantuna, E. and Kawashima, Y. and Lin, W.-H.
                   and Omanovic, D. and Piccinelli, S. and Robledo-Moreno, J.
                   and Saki, A. A. and Shee, J. and Shin, S. and Tran, M. C.
                   and Ueda, K. and Zhang, H. and Motta, M.},
  title         = {{ffsim}: Faster simulation of fermionic quantum circuits},
  journal       = {arXiv},
  year          = {2026},
  volume        = {2605.03123},
  eid           = {2605.03123},
  eprint        = {2605.03123},
  archiveprefix = {arXiv},
  doi           = {10.48550/arXiv.2605.03123},
  url           = {https://arxiv.org/abs/2605.03123}
}

@misc{ffsim_software,
  author       = {{ffsim developers}},
  title        = {{ffsim}: Fermionic quantum-circuit simulation software},
  year         = {2026},
  howpublished = {Python Package Index},
  url          = {https://pypi.org/project/ffsim/},
  note         = {Accessed 17 August 2026}
}

@article{cordella2004vf2,
  author  = {Cordella, Luigi P. and Foggia, Pasquale and Sansone, Carlo and Vento, Mario},
  title   = {A (sub)graph isomorphism algorithm for matching large graphs},
  journal = {IEEE Transactions on Pattern Analysis and Machine Intelligence},
  volume  = {26},
  number  = {10},
  pages   = {1367--1372},
  year    = {2004},
  doi     = {10.1109/TPAMI.2004.75},
  url     = {https://doi.org/10.1109/TPAMI.2004.75},
  note    = {\url{https://doi.org/10.1109/TPAMI.2004.75}}
}

@misc{jcahpc_miyabi,
  author       = {{Joint Center for Advanced High Performance Computing}},
  title        = {Supercomputer {Miyabi}},
  year         = {2026},
  howpublished = {JCAHPC},
  url          = {https://www.ccs.tsukuba.ac.jp/eng/supercomputers/},
  note         = {Accessed 17 August 2026}
}

@misc{riken_roquo,
  author       = {{RIKEN Center for Computational Science}},
  title        = {New supercomputer {ROQUO} for the quantum--{HPC} hybrid platform},
  year         = {2026},
  howpublished = {RIKEN Center for Computational Science},
  url          = {https://www.r-ccs.riken.jp/en/outreach/topics/20260619-2/index.html},
  note         = {Accessed 17 August 2026}
}

@dataset{zenodo_dataset,
  author       = {Kumar, Kunal and Horii, Hiroshi and Kobayashi, Yuri and Wakizaka, Ryo and
                  Motta, Mario and Kim, Boseong and Shirakawa, Tomonori and Yunoki, Seiji and
                  Sato, Mitsuhisa and Hamamura, Ikko},
  title        = {Reproducibility data for ``{Broken-Symmetry} {Reference} {States} for
                  {Sample-Based} {Quantum} {Diagonalization} with {Near-DMRG} {Accuracy}'':
                  {FCIDUMPs} and quantum hardware sample pools for
                  [2{Fe-2S}] and [4{Fe-4S}] clusters},
  year         = {2026},
  publisher    = {Zenodo},
  doi          = {10.5281/zenodo.23064186},
  url          = {https://doi.org/10.5281/zenodo.23064186}
}

@article{zhai2026femoco,
  author        = {Zhai, H. and Li, C. and Zhang, X. and Li, Z.
                   and Lee, S. and Chan, G. K.-L.},
  title         = {Classical computational simulation of the {FeMo} cofactor
                   model to chemical accuracy and its implications},
  journal       = {arXiv},
  year          = {2026},
  volume        = {2601.04621},
  eid           = {2601.04621},
  eprint        = {2601.04621},
  archiveprefix = {arXiv},
  doi           = {10.48550/arXiv.2601.04621},
  url           = {https://arxiv.org/abs/2601.04621}
}

@article{kreplin2020mcscf,
  author  = {Kreplin, David A. and Knowles, Peter J. and Werner, Hans-Joachim},
  title   = {{MCSCF} optimization revisited. {II}. {C}ombined first- and
             second-order orbital optimization for large molecules},
  journal = {The Journal of Chemical Physics},
  volume  = {152},
  number  = {7},
  eid     = {074102},
  year    = {2020},
  doi     = {10.1063/1.5142241},
  url     = {https://doi.org/10.1063/1.5142241},
  note    = {\url{https://doi.org/10.1063/1.5142241}}
}

@article{holmes2016hci,
  author  = {Holmes, Adam A. and Tubman, Norm M. and Umrigar, C. J.},
  title   = {Heat-Bath Configuration Interaction: An Efficient Selected
             Configuration Interaction Algorithm Inspired by Heat-Bath Sampling},
  journal = {J. Chem. Theory Comput.},
  year    = {2016},
  volume  = {12},
  number  = {8},
  pages   = {3674--3680},
  doi     = {10.1021/acs.jctc.6b00407},
  eprint  = {1606.07453},
  archiveprefix = {arXiv},
  url     = {https://doi.org/10.1021/acs.jctc.6b00407}
}

\end{document}